\magnification=\magstephalf
\baselineskip=16pt
\parskip=8pt
\rightskip=0.5truecm

\def\a{\alpha}
\def\b{\beta}
\def\d{\delta}
\def\e{\epsilon}

\def\s{\sigma}

\def\x{\xi}

\def\D{\Delta}
\def\L{\Lambda}

\def\O{\Omega}

\def\del #1{\frac{\partial^{#1}}{\partial\l^{#1}}}

\def\Const{Const\,}
\def\const{const\,}

\def\1{1}

\def\E{I\kern-.25em{E}}
\def\N{I\kern-.22em{N}}
\def\M{I\kern-.22em{M}}
\def\R{I\kern-.22em{R}}
\def\Q{I\kern-.22em{Q}}
\def\Z{{Z\kern-.5em{Z}}}

\def\C{C\kern-.75em{C}}
\def\P{I\kern-.25em{P}}

\def\del{\partial}


\def\DD{{\cal D}}
\def\EE{{\cal E}}
\def\FF{{\cal F}}

\def\HH{{\cal H}}

\def\chap #1{\line{\ch #1\hfill}}

\def\one{\hbox{J}\kern-.2em\hbox{I}}
\def\un #1{\underline{#1}}
\def\ov #1{\overline{#1}}
\def\ba{{\backslash}}
\def\sb{{\subset}}
\def\sp{{\supset}}

\def\em{{\emptyset}}


\newcount\foot
\foot=1
\def\note#1{\footnote{${}^{\number\foot}$}{\ftn #1}\advance\foot by 1}
\def\tag #1{\eqno{\hbox{\rm(#1)}}}
\def\frac#1#2{{#1\over #2}}
\def\text#1{\quad{\hbox{#1}}\quad}

\def\proposition #1{\noindent{\thbf Proposition #1: }}
\def\datei #1{\headline{\rm \the\day.\the\month.\the\year{}\hfill{#1.tex}}}

\def\theo #1{\noindent{\thbf Theorem #1: }}
\def\lemma #1{\noindent{\thbf Lemma #1: }}

\def\proof{{\noindent\pr Proof: }}
\def\proofof #1{{\noindent\pr Proof of #1: }}
\def\endproof{$\diamondsuit$}
\def\remark{{\bf Remark: }}

\def\endproof{$\diamondsuit$}

\font\pr=cmbxsl10 scaled\magstephalf
\font\thbf=cmbxsl10 scaled\magstephalf

\long\def\fussnote#1#2{{\baselineskip=10pt
\setbox\strutbox=\hbox{\vrule height 7pt depth 2pt width 0pt}
\sevenrm
\footnote{#1}{#2}
}}


\font\ch=cmbx12

\font\ftn=cmr8

\font\it=cmti10
\font\bf=cmbx10
\font\srm=cmr5


\overfullrule=0pt
\font\tit=cmbx12
\font\aut=cmbx12
\font\aff=cmsl12
{$  $}
\vskip0.8truecm
\centerline{\tit 
WEAKLY GIBBSIAN REPRESENTATIONS FOR JOINT MEASURES
}
\vskip.2truecm
\centerline{\tit OF QUENCHED LATTICE SPIN MODELS
\footnote{${}^*$}{\ftn Work
supported by the DFG
Schwerpunkt `Wechselwirkende stochastische Systeme hoher Komplexit\"at'
}}
\vskip1.7truecm
\vskip.3truecm
\centerline{\aut  Christof K\"ulske\footnote{${}^{1}$}{\ftn
e-mail: kuelske@wias-berlin.de}
}


\vskip.1truecm
\centerline{\aff WIAS}
\centerline{\aff Mohrenstrasse 39}
\centerline{\aff D-10117 Berlin, Germany}
\vskip0.6truecm\rm

\noindent {\bf Abstract:}
Can the joint measures of quenched disordered lattice 
spin models (with finite range) 
on the product of spin-space and disorder-space
be represented as (suitably generalized) Gibbs measures of an 
``annealed system''? -
We prove that there is always a potential (depending on both spin 
and disorder variables) that converges absolutely on a set 
of full measure w.r.t. the joint measure (``weak Gibbsianness'').    
This ``positive'' result is surprising when contrasted with 
the results of a previous paper [K6], where we investigated the measure 
of the set of discontinuity points of the conditional 
expectations (investigation of ``a.s. Gibbsianness'').  
In particular we gave natural ``negative'' examples 
where this set is even of measure one
(including the random field Ising model). 

Further we discuss conditions giving 
the convergence of vacuum potentials 
and conditions for the decay of the joint potential 
in terms of the decay of the disorder average 
over certain quenched correlations. We apply 
them to various examples. From this one
typically expects the existence of a potential 
that decays superpolynomially outside a set 
of measure zero. 
Our proof uses a martingale argument
that allows to cut (an infinite volume analogue of)  
the quenched free energy into local pieces, along with 
generalizations of Kozlov's constructions. 



\vfill
     ${}$
\eject



\magnification=\magstephalf
\baselineskip=16pt
\parskip=8pt
\rightskip=0.5truecm

\def\a{\alpha}
\def\b{\beta}
\def\d{\delta}
\def\e{\epsilon}

\def\s{\sigma}

\def\x{\xi}

\def\D{\Delta}
\def\L{\Lambda}

\def\O{\Omega}

\def\del #1{\frac{\partial^{#1}}{\partial\l^{#1}}}

\def\Const{Const\,}
\def\const{const\,}

\def\1{1}

\def\E{I\kern-.25em{E}}
\def\N{I\kern-.22em{N}}
\def\M{I\kern-.22em{M}}
\def\R{I\kern-.22em{R}}
\def\Z{{Z\kern-.5em{Z}}}

\def\C{C\kern-.75em{C}}
\def\P{I\kern-.25em{P}}

\def\del{\partial}


\def\DD{{\cal D}}
\def\EE{{\cal E}}
\def\FF{{\cal F}}

\def\HH{{\cal H}}

\def\chap #1{\line{\ch #1\hfill}}

\def\one{\hbox{J}\kern-.2em\hbox{I}}
\def\un #1{\underline{#1}}
\def\ov #1{\overline{#1}}
\def\ba{{\backslash}}
\def\sb{{\subset}}
\def\sp{{\supset}}

\def\em{{\emptyset}}


\newcount\foot
\foot=1
\def\note#1{\footnote{${}^{\number\foot}$}{\ftn #1}\advance\foot by 1}
\def\tag #1{\eqno{\hbox{\rm(#1)}}}
\def\frac#1#2{{#1\over #2}}
\def\text#1{\quad{\hbox{#1}}\quad}

\def\proposition #1{\noindent{\thbf Proposition #1: }}
\def\datei #1{\headline{\rm \the\day.\the\month.\the\year{}\hfill{#1.tex}}}

\def\theo #1{\noindent{\thbf Theorem #1: }}
\def\lemma #1{\noindent{\thbf Lemma #1: }}

\def\proof{{\noindent\pr Proof: }}
\def\proofof #1{{\noindent\pr Proof of #1: }}
\def\endproof{$\diamondsuit$}
\def\remark{{\bf Remark: }}

\def\endproof{$\diamondsuit$}

\font\pr=cmbxsl10 scaled\magstephalf
\font\thbf=cmbxsl10 scaled\magstephalf

\long\def\fussnote#1#2{{\baselineskip=10pt
\setbox\strutbox=\hbox{\vrule height 7pt depth 2pt width 0pt}
\sevenrm
\footnote{#1}{#2}
}}


\font\ch=cmbx12

\font\ftn=cmr8

\font\it=cmti10
\font\bf=cmbx10
\font\srm=cmr5




\chap{I. Introduction\hfill}

Consider the joint measure corresponding 
to a random infinite volume Gibbs measure of a disordered 
lattice spin system. 
By this we mean the measure $\P(d\eta)\mu[\eta](d\s)$ 
on the product space of disorder variables 
$\eta$ and spin variables $\s$.
Here $\mu[\eta](d\s)$ is a random Gibbs measure
and $\P$ is the a-priori distribution of the disorder variables.  
Prototypical examples for such quenched random systems
are the random field Ising model or an 
Ising model with random couplings. 

In this paper we investigate of the question: 
When can these measures be understood as
Gibbs measures on the skew space, respectively suitable 
generalizations thereof? More specifically, are there 
well-defined Hamiltonians, given in terms of 
interaction potentials depending on both spin and disorder 
variables, that provide an annealed description for such a system?
The formal description of disordered systems in terms 
of such potentials was termed ``Morita's 
equilibrium ensemble approach to disordered systems'' 
(see e.g [Ku1,2], [MKu], [Mo], [SW] and references in [Ku2]) 
in the theoretical physics community. 
However, the existence of such Hamiltonians was never investigated 
rigorously but taken for granted, and various approximation schemes 
were based on the truncation of the corresponding potentials.
In this respect there is an analogy between the problems of the existence 
of joint potentials and of the existence
of ``renormalized potentials'' that are supposed to  
give a Gibbsian description of a measure that appears as 
an image measure of a Gibbs measure under a renormalization 
group transformation. There is a huge literature 
about the latter ones but the present question has 
remained mathematically neglected until recently ([EMSS], [K6]).  

Now, mathematically, it turns out that the answer to our question 
is a somewhat complicated but interesting one. 
It depends on the kind of generalization of 
the notion of Gibbsianness 
one is asking for and on the specific system. 
Therefore such joint measures corresponding
to quenched random systems provide a rich class of 
examples to illustrate the subtleties of 
the different generalizations of the notion of Gibbsianity. 
We believe that, while interesting in itself,  
the study of these measures 
is also valuable for the understanding of the fine (and not 
always very intuitive) distinctions that are necessary 
if one attempts to extend Gibbsian theory to 
non-Gibbsian measures.
   
Recall that Gibbs measures of an infinite volume  
lattice system are characterized by the fact 
that their conditional expectations (given 
the values of the variables outside of a finite 
volume) can be written in terms of an 
absolutely convergent interaction potential. 
Equivalently, they are the measures for which 
these conditional expectations are 
continuous functions of the conditioning.
(The less trivial part of the equivalence, i.e. 
existence of a potential assuming continuity 
of conditional expectations,   
is due to  the construction of [Koz]). 
For general information about 
scenarios of the failure of the 
Gibbsian property for lattice measures and 
possible generalizations of Gibbsianness 
see e.g. [F],[E],[DS],[BKL],[MRM], [MRSM], references therein,
and the basic paper [EFS]. 

In the first mathematical paper [EMSS] which studied a joint 
measure of a quenched random system  
it was shown that the joint measure resulting from the
diluted Ising ferromagnet at low temperatures 
is not a Gibbs measure in the strict sense described above:
[EMSS] showed that there is a point of 
essential discontinuity in the conditional expectations as a function 
of the conditioning. So, the measure does not 
allow for a  Hamiltonian constructed from  
an absolutely summable interaction potential. 
However, the set of such discontinuities has zero measure
in this example. Measures with this property
are commonly called
``almost Gibbsian'' measures. The notion
of ``almost Gibbsianness''
is a  straightforward measure-theoretic attempt 
to generalize the classical notion of 
Gibbsianness where the conditional expectations
are continuous everywhere.  
 
In a recent paper [K6] we investigated the question of 
discontinuity of the conditional expectations in the general
setup of quenched lattice spin 
systems with finite range quenched 
Hamiltonians depending on independent disorder variables. 
In particular, we gave an example where the set of 
discontinuities was even a {\bf full measure set}. 
So, even worse, this measure even fails to be ``almost Gibbsian''! 
The example was the random field Ising model in the 
phase transition regime.  
It is particularly illuminating because 
it shows in a transparent manner a more general fact:
The question of discontinuity of the conditional
expectations is related to whether a discontinuity  
can be felt on certain local expectations of the 
quenched measure by varying the disorder variables arbitrarily 
far away. The local expectation under consideration 
is just the magnetization for the random field Ising model; 
more generally this has to be replaced by the spin-observable
conjugate to the independent disorder variables.
In [K6] we also discussed another interesting phenomenon: 
We argued that whether the set of discontinuity points is of measure 
zero or one can depend on the random Gibbs measure, 
for the same choice of the parameters. 
This phenomenon should appear in the 
random bond ferromagnet at low temperatures, weak disorder, 
and high dimensions: We argued that it is to be expected 
that the set of discontinuities 
should be of measure zero for the ferromagnetic 
plus state while it should be of measure one 
for the random Dobrushin state.

While we focused on ``almost Gibbsianness'' in [K6], 
the aim of the present paper is to find out what 
can be said about ``weak Gibbsianness''.  
The latter notion is a different attempt to weaken 
(even more) the classical notion of Gibbs measure. Here one 
requires only the existence of a potential that 
is convergent (or even absolutely convergent) 
on a full measure set (and not 
necessarily everywhere). [MRM] noted 
that, in general, an almost Gibbsian measure always has 
a potential that is convergent 
on a set of full measure. It is however {\bf not} 
expected that there is always an {\bf absolutely} convergent potential
in this situation.  
Also, [MRM] gave an example of a measure 
having a convergent potential which was not almost Gibbsian.

In this note we will give a completely general positive 
answer to the question of weak Gibbsianness for our measures. 
That is, at least from the point of view of weak Gibbsianness, 
the situation gets easier again. 
We will show:

{\bf The joint measures corresponding to a random infinite volume 
Gibbs measure always posses a potential that converges absolutely 
on a full measure set.} 

For the specific example of the random field Ising model 
in the phase transition regime this gives, together 
with the result of [K6] the following interesting statement: 
The set of discontinuity points of the joint measure 
has full measure, but still there is a potential 
that converges absolutely on a set of full measure. 
\footnote{$^1$}{Recently [Le] constructed an independent 
example  of a lattice measure (not related to random systems)
to illustrate that this phenomenon can really occur.   
}
So, almost Gibbsianness does not hold,
but weak Gibbsianness does (even in a strong form).  
In fact, we expect the convergence to be very fast on a 
set of measure one (see Chapter V.)

Our existence result  
is true for any quenched lattice spin systems with finite range quenched 
Hamiltonians depending on sitewise independent disorder variables.
We stress here that {\bf no continuity assumptions} at all
are needed on the measures involved. This may seem surprising
and is a main non-trivial point. 
Let us describe our results at first in words, 
before we put them down in precise formulas.  
They will all have the following form: 
We construct a potential and 
explain its properties and 
how it is related to the given ``quenched potential''
that is the starting point and 
defines the system we are dealing with. 

Now, to put the first result in perspective, 
we remark that in the case of a general lattice measure, 
the existence of an a.s. convergent potential 
can be obtained once there is at least one direction of (a.s.) continuity
for the conditional expectations 
(see [MRM]) using the corresponding vacuum potential.  
Due to the special form of the joint measures we are considering
here, we can improve on this in our case (see Theorems 2.1,2.3). 
For this we take advantage of the specific form of the 
infinite volume conditional expectations of the joint 
measures derived in Chapter II.  
The trick to get the stronger result is to use not a 
vacuum potential, but a different one; 
this will allow to conclude convergence of the potential 
by a soft martingale argument. 
From this we can get an existence result for an a.s. 
{\bf absolutely} convergent potential generalizing 
the one of [Koz]. 
We remark that also for this latter step we are again 
exploiting the special nature of our measures; it would 
not work for a general lattice measure. 

Nevertheless, 
it is also interesting to see what can be said about 
the convergence of vacuum potentials (see Theorem 2.2). 
For this we need in fact some continuity, 
conveniently expressed 
in terms of the behavior of the corresponding 
infinite volume Gibbs state: 
One needs continuity of the corresponding 
infinite volume quenched Gibbs-expectation  
of the spin-observable conjugate to the independent disorder variables, 
as a function of the quenched variables, 
in the direction of a 
certain realization of the disorder.  
These are the same observables whose behavior was crucial also for the 
question of ``almost sure Gibbsianness''. 

Next, if one would like to have more information 
about the decay of the potential, one has 
to assume some information about the clustering 
properties of the quenched random system.
We relate the decay of a joint potential
to the decay of disorder-averages
of certain quenched correlations in Theorem 2.4. 
These  correlations are taken between the spin-
observables conjugate to the independent disorder variables, 
the same ones as above. 
Physically, superpolynomial decay of such 
averaged correlations is typically to be expected (off the 
critical point).  
So, we should typically expect the existence 
of a potential that decays superpolynomially 
outside of a set of measure zero. 
Of course, to prove it, specific analysis of the system 
under consideration is needed, which can be very hard. 

The paper is organized as follows. 
In Chapter II we define the class of models 
we will treat and state our results in precise terms. 
In Chapter III we prove the important formula for the 
infinite volume conditional expectations of the joint measure 
that is the starting point of the following. 
In Chapter IV we will prove the theorems stated in Chapter II. 
In Chapter V we will discuss the examples 
of the random field Ising model, Ising models with 
random couplings (which also fit into our framework), 
and the diluted Ising ferromagnet, including some heuristic considerations. 
\bigskip
\bigskip

\chap{Acknowledgments:}

The author thanks A.C.D.van Enter for pointing out the 
physical relevance of the problem and various   
comments on an earlier draft of the paper. 

\bigskip
\bigskip

\chap{II. The Models and the Results\hfill}

Denote by $\O=\O_{0}^{\Z^d}$
the space of {\bf spin-configurations} 
$\s=\left(\s_x \right)_{x\in \Z^d}$, 
where $\O_0$ is a finite set. 
Similarly we denote by $\HH=\HH_{0}^{\Z^d}$ 
the space of {\bf disorder variables}
$\eta=\left(\eta_x \right)_{x\in \Z^d}$
entering the model, where $\HH_0$ is a finite set.
Each copy of $\HH_0$ carries a measure $\nu(d\eta_x)$
and $\HH$ carries the product-measure over the sites,
$\P=\nu^{\otimes_{\Z^d}}$. We denote the 
corresponding expectation by $\E$. 
The space of joint configurations $\bar \O:=\O\times\HH=
\left(\O_{0}\times \HH_{0} \right)^{\Z^d}$ 
is called {\bf skew space}. It is equipped with
the product topology and the corresponding Borel sigma algebra. 

A {\bf potential on the joint 
variables} is a family $U$ of real functions 
$U_{A}:\bar \O\rightarrow \R$ where  $A$ runs over 
the subsets of $\Z^d$ s.t. 
$U_{A}(\x)$ depends only on $\x_A$. 
We consider disordered models whose finite 
volume Gibbs-measures can be written in  
terms of a potential $\Phi=(\Phi_A)_{A\sb \Z^d}$
on the joint variables. 
In this context 
we will call $\Phi$ the {\bf disordered potential}.
We fix a realization of the disorder $\eta$ and 
define probability measures 
$\mu_{\L}^{\s^{\hbox{\srm b.c.}}}[\eta]$ on the spin space 
$\O$, called the {\bf quenched finite volume Gibbs measures}, by 
$$
\eqalign{
&\mu_{\L}^{\s^{\hbox{\srm b.c.}}}[\eta](\s)
:=\frac{
e^{-\sum_{A:A\cap \L\neq \em}
\Phi_{A}( \s_{\L}\s^{\hbox{\srm b.c.}}_{\Z^d\ba \L},
\eta) 
}
}{
\sum_{\tilde\s_{\L}}
e^{-\sum_{A:A\cap \L\neq \em}
\Phi_{A}( \tilde\s_{\L}\s^{\hbox{\srm b.c.}}_{\Z^d\ba \L},
\eta)
}
} 1_{\s_{\Z^d\ba \L}=\s^{\hbox{\srm b.c.}}_{\Z^d\ba \L}}
}
\tag{2.1}
$$
The finite-volume 
summation is over $\s_{\L}\in \O_{0}^{\L}$. 
The symbol $\s_{\L}\s^{\hbox{\srm b.c.}}_{\Z^d\ba \L}$
denotes the configuration in $\O$ that is given 
by $\s_{x}$ for $x\in\L$ and by
$\s^{\hbox{\srm b.c.}}_{x}$ for $x\in \Z^d\ba \L$.
We assume for simplicity {\bf finite range}, i.e. 
that $\Phi_{A}=0$ for $\hbox{diam}A>r$. 
This form is really quite general.
It is a simple matter to write the random field 
Ising model or an Ising model with disordered nearest 
neighbor couplings in the above form. 

Next, we suppose from the beginning 
that we have the existence of a weak limit 
$$
\eqalign{
&\lim_{\L\uparrow\Z^d}
\mu^{\s^{\hbox{\srm b.c.}}_{\del\L}}_{\L}[\eta]
= \mu[\eta]
}
\tag{2.2}
$$
for $\P$-a.e. $\eta\equiv \eta_{\Z^d}$ 
with a nonrandom boundary condition $\s^{\hbox{\srm b.c.}}$.  
In ferromagnetic examples like the random field Ising 
model this can be concluded by monotonicity arguments. Note 
that there is however no general argument that would give 
the existence of this limit - indeed it is expected 
to fail e.g. for low temperature spinglasses.\footnote{$^2$}
{Side-remark about the relation to ``metastates'': 
It is this existence problem that led to the introduction 
of the general notion of metastates, which are distributions 
of Gibbs-measures, see e.g. [NS1]-[NS5], [K2]-[K5].
Also, more generally than in the present note, 
in large parts of [K6] we did not assume the a.s. convergence 
of the random finite volume Gibbs measures, but only the weaker
property of convergence of the corresponding finite volume 
joint measures. Assuming the existence of a corresponding 
metastate, such a measure $K$ is its barycenter. 
The case of the present note corresponds to the trivial metastate 
which is supported only on a single state $\mu[\eta]$. 
}

Assuming (2.2) 
it follows that $\mu_{\infty}[\eta_{\Z^d}]$
is an infinite-volume Gibbs measure for $P$-a.e. $\eta$
that depends measurably on $\eta$.
We look at spins and disorder variables at the same
time and define {\bf joint spin variables}
$\xi_x=(\s_x,\eta_x)\in \O_{0}\times \HH_{0}$. 
The central object of our study 
is the corresponding 
{\bf infinite volume joint measure}
on the skew space $\left(\O_{0}\times \HH_{0} \right)^{\Z^d}$
defined by 
$$
\eqalign{
&K(d\s,d\eta):=\P(d\eta) \mu[\eta](d\s)
}
\tag{2.3}
$$
We say that a potential $U$ on the joint 
variables is a {\bf potential for the joint measure} $K$ if
$U$ produces the correct conditional expectations for $K$, i.e.  
$$
\eqalign{
&\frac{e^{-\sum_{A:A\cap \L\neq \em}U_A(\xi)}
}{
\sum_{\tilde \x_{\L}}
e^{-\sum_{A:A\cap \L\neq \em}U_A(\tilde\xi_{\L}\xi_{\Z^d\ba \L})}
}
=K[\xi_{\L}|\xi_{\Z^d\ba \L}]
}
\tag{2.4}
$$
for $K$-a.e. $\xi$.
This work is about the existence of such a potential. 
It provides a description of the joint measure 
as an ``annealed system''. This notion should not 
be confused with the ``trivial'' annealed system 
appearing in the next definition. 

We call a potential $U^{\srm{ann}}$ on the joint 
variables a {\bf potential for the annealed system}
if it is finite range and 
produces the annealed local specification, i.e. 
$$
\eqalign{
&\frac{e^{-\sum_{A\cap \L\neq \em}
U^{\srm{ann}}_{A}(\s_{\L}\s^{\hbox{\srm b.c.}}_{\Z^d\ba \L},
\eta_{\L}\eta^{\hbox{\srm b.c.}}_{\Z^d\ba \L})}}{
\sum_{\tilde\s_{\L},\tilde\eta_{\L}}
e^{-\sum_{A\cap \L\neq \em}
U^{\srm{ann}}_{A}( \tilde\s_{\L}\s^{\hbox{\srm b.c.}}_{\Z^d\ba \L},
\tilde\eta_{\L}\eta^{\hbox{\srm b.c.}}_{\Z^d\ba \L})
}
}
=\frac{\nu(\eta_{\L})e^{-\sum_{A\cap \L\neq \em}
\Phi_{A}(\s_{\L}\s^{\hbox{\srm b.c.}}_{\Z^d\ba \L},
\eta_{\L}\eta^{\hbox{\srm b.c.}}_{\Z^d\ba \L})}}{
\sum_{\tilde\s_{\L},\tilde\eta_{\L}}
\nu(\tilde\eta_{\L})e^{-\sum_{A\cap \L\neq \em}
\Phi_{A}( \tilde\s_{\L}\s^{\hbox{\srm b.c.}}_{\Z^d\ba \L},
\tilde\eta_{\L}\eta^{\hbox{\srm b.c.}}_{\Z^d\ba \L})
}
}
} 
\tag{2.5}
$$
We call this system ``annealed'' because the r.h.s. describes 
a joint system given by an Hamiltonian which is simply 
the quenched Hamiltonian 
and a priori measure given by the independent distribution 
$\P$ for the disorder variables. Of course, its  
properties may differ completely from the quenched system. 
Trivially, one such potential is $U^{\srm{ann}}_{A}(\s,\eta)=
\Phi_{A}(\s,\eta)-1_{A=\{x\}}\log \nu(\eta_x)$.
We remark that, of course, the problem of classifying the 
equivalent potentials $U$ for given $\nu, \Phi$ 
is long solved and can 
be found in [Geo], see paragraphs (2.3) and (2.4) therein.

Finally, a potential $U$ is called {\bf summable for $\x$} 
if, for any $\L\sb \Z^d$, we have that the limit   
$\lim_{\D\uparrow\Z^d}\sum_{A:A\cap \L\neq \em,A\sb \D}U_A(\xi)
=:\sum_{A:A\cap \L\neq \em}U_A(\xi)$
exists and is independent of the sequence of $\D$'s.
This is needed for the sums in (2.4) to make sense. 
$U$ is called {\bf absolutely summable for $\x$} 
if, for any $\L\sb \Z^d$ we have that
$\sup_{\D\sb\Z^d}\sum_{A:A\cap \L\neq \em,A\sb \D}\left|
U_A(\xi)\right|< \infty$. 

Now, the most natural approach 
to find a potential for the joint measure is to 
write down a formal vacuum potential on the joint 
space and ask what we can say about its convergence
(see Theorem 2.2).
We remind the reader that a potential $U$ is called vacuum 
potential with vacuum $\hat\x$, 
if $U_{A}(\xi_{A\ba x}\hat \x_x)  = 0$ whenever $x\in A$.
However, it turns out that we get our strongest 
general existence result of Theorem 2.1 for a different potential.   
To this end, let $\a(d\x)$ be a product probability measure.  
Then, a potential $U$ is called {\bf $\a$-normalized} if 
$\int\a_x(d\tilde\x_x) 
U_{A}(\xi_{A\ba x}\tilde \x_x)   
= 0$ whenever $x\in A$. Obviously, for $\a=\d_{\hat\xi}$, 
an $\a$-normalized potential is a vacuum potential with 
vacuum $\hat \xi$. This notion was first introduced by Israel [I]
but we use the terminology of Georgii.   

\bigskip
In the following we assume that we are given a joint 
measure of the type (2.5)
corresponding to a quenched random lattice model defined
by (2.1), (2.2). Then the following statements hold. 

\theo{2.1 (Existence of a.s. summable potential)}{\it There exists a 
potential $U$ for $K$ that is summable 
for $K$-a.e. $\xi$. This is 
true under no further assumptions on 
the continuity properties of $\mu[\eta]$.
This potential has the form 
$U(\s,\eta)=U^{\srm{ann}}(\s,\eta)+
U^{\hbox{\srm{fe}}}_{\mu}(\eta)$. 
In this equation $U^{\srm{ann}}$ is 
any finite range potential for the annealed system, independently 
chosen of the second term.  

$U^{\hbox{\srm{fe}}}_{\mu}$ is a potential depending only on $\eta$
which is convergent for $\P$-a.e. $\eta$. 
As a potential on the disorder space it is $\P$-normalized. 
In general, 
two different measurable infinite volume Gibbs-states 
$\mu:\eta\mapsto\mu[\eta]$ 
corresponding to the same random local specification
will yield different $U^{\hbox{\srm{fe}}}_{\mu}$.
}

The notation $U^{\hbox{\bf\srm{fe}}}_{\mu}(\eta)$ 
is meant to suggest to the reader, that this potential
comes from a decomposition into local terms
of what in finite volume 
would be the disorder dependent
free energies of the quenched system.  
This will become clear in the proofs. 
An analogous finite volume 
quantity is called ``disorder potential'' in [Ku2].

To describe the kind of continuity we need for 
the existence of the vacuum potential in detail
we need some more notation.  
For a subset $V\sb \Z^d$, we call the expression 

$$
\eqalign{
&\D H_{V}(\s_{\ov{V}},\eta^1_{V},\eta^2_{V},\eta_{\del {V}})
:= \sum_{A: A\cap V\neq \em}\Bigl(\Phi_{A}\left(
\s_{\ov{V}},\eta^1_{V}\eta_{\del {V}}\right) 
-\Phi_{A}\left(
\s_{\ov{V}},\eta^2_{V}\eta_{\del {V}}
\right)
\Bigr)
\cr
}
\tag{2.6}
$$
the 
{\bf $V$-variation of the Hamiltonian w.r.t.  
the disorder variables}. 
To denote the corresponding function on the spin-variables
obtained by fixing the disorder variables 
we will drop the spin-variable $\s$ on the l.h.s. of (2.6). 
In particular, for $V=\{x\}$, the expression (2.6)
is the observable conjugate to the independent disorder 
variable $\eta_x$. We put 
$$
\eqalign{
&Q_{x}(\eta^{1}_{x},\eta^{2}_{x},\eta_{\Z^d\ba x})
:=\mu[\eta^2_{x},\eta_{\Z^d\ba x}]
(e^{-\D H_{x}(\eta^1_{x},\eta^2_{x},\eta_{\del {x}})})
\cr
}
\tag{2.7}
$$
for its  quenched expectation.

\theo{2.2 (A.s. summability of vacuum potential)}
{\it Suppose moreover that there exists 
a direction $\hat\eta$ of a.s. continuity for the quenched 
expectation of the spin observable conjugate 
to the disorder variables, i.e.  
$$
\eqalign{
&\lim_{\L\uparrow\Z^d}Q_
{x}(\eta^{1}_{x},\eta^{2}_{x},
\eta_{\L\ba x}\hat\eta_{\Z^d\ba \L})
=Q_{x}(\eta^{1}_{x},\eta^{2}_{x},
\eta_{\Z^d\ba x})
\cr
}
\tag{2.8}
$$
for all $x$, $\eta^1_x$, $\eta^2_x$, for $\P$-a.e. $\eta$.
We assume that $Q$ is defined by the weak limit (2.2) 
and (2.7) and this weak limit exists for $\P$-a.e. $\eta$. 
Here we have fixed a nonrandom boundary condition 
$\s^{\hbox{\srm b.c.}}$ for those $\eta$ that are {\bf not} 
in the $\P$-zero-set  of $\eta$'s of the form 
$(\eta_{\L}\hat\eta_{\Z^d\ba \L})$. Moreover we assume 
that (2.2) also exists for $\hat\eta$ (and thus 
for all the countably many $\eta$'s of the form  
$(\eta_{\L}\hat\eta_{\Z^d\ba \L})$), with some 
{\bf possibly different}  boundary condition $\hat\s^{\hbox{\srm b.c.}}$.

Then there is a vacuum potential 
$V^{\hbox{\srm{fe}}}_{\mu}(\eta)$ on the disorder space
with vacuum $\hat \eta$ s.t. 
$U'(\s,\eta)=U^{\srm{ann}}(\s,\eta)+V^{\hbox{\srm{fe}}}_{\mu}(\eta)$
is a potential for the joint measure $K$ 
which is summable $K$-a.s.. Here
$U^{\srm{ann}}$ is any arbitrarily chosen finite
range potential for the annealed system.}

Note that our hypothesis is weaker than
requiring a.s. continuity of $\mu[\eta]$ itself 
in direction $\hat\eta$ 
(by which one understands continuity of {\bf all} probabilities
$\mu[\eta](\s_{\L})$ in this direction.)
Note that, in general, 
the {\bf same} choices of boundary conditions 
to construct the state $\mu[\hat\eta]$, and the 
state $\mu[\eta]$ for typical $\eta$ might 
yield a state of {\bf different} type (see V(iii)). 

Now, in the situation of Theorem 2.2, fix any $\hat \s$. Then
we can in particular choose $U^{\srm{ann}}(\s,\eta)$
to be the unique vacuum potential for 
the annealed system with vacuum $(\hat\s,\hat\eta)$. 
\footnote{$^3$}{
A clear proof of the existence of an 
$\a$-normalized convergent potential  
in the case of {\it continuous} conditional expectations
can be found in [Geo] Theorem (2.30). Under our assumptions
of discrete joint spin space and finite range of the defining
disordered potential $\Phi$ this theorem
shows in particular: For any $\a$ there exists 
a unique equivalent $\a$-normalized potential for the 
annealed system with the same range.

} This gives the simple

\noindent{\thbf Corollary 1: }
{\it If $\hat\eta$ is a direction of continuity for $\mu(\eta)$, 
for any $\hat\s\in \O$, the formal vacuum potential for $K$
with vacuum $\hat \x=(\hat\s,\hat\eta)$
is convergent for $K$-a.e $\xi$.
Here we have assumed that $\mu[\eta]$ is defined 
by the weak limit (2.2) with boundary conditions 
as in the hypothesis of Theorem 2.2.}

\noindent\remark If $K$ is translation-invariant, so are the 
potentials constructed in the proof of Theorem 2.1 and Theorem 2.2. 
In general, they need not be absolutely summable. 

The proof of Theorem 2.2 also gives

\noindent{\thbf Corollary 2: }
{\it The sum $\sum_{A:A\cap \L\neq \em}
\int\P(d\tilde\eta)V^{\hbox{\srm{fe}}}_{\mu; A}(\tilde\eta)$ converges. 
Hence 
$U^{\srm{ann}}_A(\s,\eta)+\Bigl[ V^{\hbox{\srm{fe}}}_{\mu; A}(\eta)
- \int\P(d\tilde\eta)V^{\hbox{\srm{fe}}}_{\mu; A}(\tilde\eta)\Bigr]
$ is a potential for the joint measure which is summable 
$K$-a.s., too.}\footnote{$^4$}{This proves general existence 
of potentials of the form generalizing 
the one that was written down in finite volume 
in [Ku2 (32)] for the special case of the 
dilute Ising model, where no proof of the 
infinite volume limit was given
(see also Chapter V).}

From Theorem 2.1 one can obtain an absolutely summable
potential, if one gives up translation invariance.

\theo{2.3 (Existence of a.s.  
absolutely convergent potential)}
{\it There exists an {\bf a.s. absolutely summable} potential 
$U^{\hbox{\srm abs}}$ for the joint measure $K$ of the form 
$U^{\hbox{\srm abs}}(\s,\eta)=U^{\srm{ann}}(\s,\eta)+
U^{\hbox{\srm fe, abs }}_{\mu}(\eta)$. 
Here, as above, $U^{\srm{ann}}$ is any finite range potential
for the annealed system. 
$U^{\hbox{\srm fe, abs }}_{\mu}$ is a potential depending only on 
$\eta$ which is {\bf absolutely} 
convergent for $\P$-a.e. $\eta$. 
$U^{\hbox{\srm fe, abs}}_{\mu}$ 
is not necessarily translation invariant 
even if $K$ is translation invariant. 
As in Theorem 2.1, this results holds under 
no further continuity assumptions on $\mu[\eta]$. 
}

\remark In fact the new `free energy'
potential $U^{\hbox{\srm fe, abs}}_{\mu}$ is even 
integrable w.r.t. $K$ (which is to say integrable w.r.t. $\P$).
There is no estimate on the speed of convergence. 

$U^{\hbox{\srm fe, abs }}_{\mu}(\eta)$
is supported on a very sparse system 
of subsets of $\Z^d$. It is obtained by a resummation  
of the $\P$-normalized `free energy' potential 
$U^{\hbox{\srm{fe}}}_{\mu}$ from 
the construction Kozlov used on the vacuum potential
in the case of a measure with continuous conditional
expectations [Koz].
We remark that the same construction can in general 
{\bf not} be applied to the vacuum
potential $V^{\hbox{\srm{fe}}}_{\mu}$
of Theorem 2.2, unless there is additional 
information on its decay.    

\medskip
\remark Let us also comment on the easy case, 
when $Q$ is continuous {\bf everywhere}, by which we mean that 
$$
\eqalign{
&\lim_{\L\uparrow\Z^d}\sup_{\hat\eta}\left|
Q_{x}(\eta^{1}_{x},\eta^{2}_{x},\eta_{\L\ba x}\hat\eta_{\Z^d\ba \L})
- Q_{x}(\eta^{1}_{x},\eta^{2}_{x},\eta_{\Z^d\ba x})\right|
=0
}
\tag{2.9}
$$
for all $\eta$ and all $x$, $\eta^1_x$, $\eta^2_x$. 
Then, the infinite 
volume conditional expectations of $K$ are continuous, 
and so $K$ is a Gibbs measure. 
The ``free energy potentials'' $U^{\hbox{\srm{fe}}}_{\mu}$ 
(of Theorem 2.1) and $V^{\hbox{\srm{fe}}}_{\mu}$ 
(of Theorem 2.2) are both convergent everywhere.  
Furthermore, the stronger version 
of Theorem 2.3 holds where ``a.s. absolute summability'' 
is strengthened to ``absolute summability everywhere''.

To get an absolutely summable 
potential for the {\it joint measure} that is also 
translation invariant, more information 
on the clustering properties of the quenched system 
on the average is needed. 
Theorem 2.4 below describes the existence of an a.s. absolutely summable
potential that is {\bf translation invariant}, if the measure $K$
is. Moreover it gives information about the decay 
of this potential.

\theo{2.4 (A.s. absolutely summable translation invariant potential)}{\it 

Assume that the  averaged
quenched correlations satisfy the decay property
$\sum_{m=1}^{\infty} m^{2d-1}\bar c(m)<\infty$
where $\bar c(m):=\sup_{{x,y:|x-y|=m}\atop{
\eta_x,\eta_y\in \HH_0}} 
\int\P(d\tilde \eta)
\left| c_{x,y}(\eta_x,\eta_y,\tilde \eta)\right|$ 
with
$$
\eqalign{
&c_{x,y}(\eta_x,\eta_y,\tilde \eta)\cr
&:=
\mu[\tilde\eta]
\left(e^{-\D H_{\{x,y\}}
(\eta_{\{x,y\}},\tilde\eta_{\{x,y\}},\tilde\eta
\bigl|_{\del {\{x,y\}}})}\right)-\mu[\tilde\eta]
\left(
e^{-\D H_{x}(\eta_{x},\tilde\eta_{x},\tilde\eta
\bigl|_{\del {x}})}
\right)
\mu[\tilde\eta]
\left(
e^{-\D H_{y}(\eta_{y},\tilde\eta_{y},\tilde\eta
\bigl|_{\del {y}})}
\right)\cr
}
\tag{2.10}
$$
Then there is an a.s. absolutely summable potential 
$U^{\hbox{\srm{fe,abs,inv}}}_{\mu}(\eta)$ on the disorder space
s.t. $U^{(4)}(\s,\eta)
=U^{\srm{ann}}(\s,\eta)+U^{\hbox{\srm{fe,abs,inv}}}_{\mu}
(\eta)$
is a potential for the joint measure $K$, 
for any arbitrarily chosen finite range potential 
$U^{\srm{ann}}$ for the annealed system.

If $K$ is translation invariant, then 
$U^{\hbox{\srm{fe,abs,inv}}}_{\mu}(\eta)$ is translation 
invariant, too. 
}

\remark Again, the potential is even integrable. 
Moreover, for 
any nonnegative translation invariant 
function $w(A)$ giving weight  
to a subset $A\sb \Z^d$ we have the following
estimate on its decay
$$
\eqalign{
&\sum_{A:A\ni x_0} w(A)
\int\P(d\tilde\eta)
\left |
U^{\hbox{\srm{fe,abs,inv}}}_{\mu; A}(\tilde \eta)\right|
\leq C_1+C_2\sum_{m=2}^{\infty} m^{2d-1} \bar w(m) \bar c(m)
\cr
}
\tag{2.11}
$$
where $\bar w(m):= w\left(\{z\in \Z^d;z\geq  0, |z|\leq m \}\right)$
where $\geq $ denotes the lexicographic order.
The constants $C_1$, $C_2$ are related to a-priori 
bounds on $\D H_x$. 

Under the stronger condition that we 
have bounds of the same  form on the 
$\sup_{{x,y:|x-y|=m}\atop{
\eta_x,\eta_y\in \HH_0}}$ 
$\sup_{\tilde \eta}
\left| c_{x,y}(\eta_x,\eta_y,\tilde \eta)\right|$ 
the absolute convergence is not only a.s. but everywhere,  
and (2.11) holds for all realizations 
without the $\P$-integral (with non-random constants).

\bigskip\bigskip
\bigskip\bigskip

\chap{III. The infinite volume conditional expectations\hfill}

We start with a suitable representation of 
the infinite volume conditional expectations of the 
joint measure. 

We write $\x=(\s,\eta)$ here and below, so that, for any 
set $A\sb \Z^d$ we have $\x_A=(\s_A,\eta_A)$. 
Recall that $r$ is the range of the defining potential $\Phi$.
We write $\ov {A}=\{y\in \Z^d, d(y,A)\leq r\}$ for 
the $r$-neighborhood  of a set $A$, and put $\del A=\ov{A}\ba A$.

\proposition{3.1}{\it  Assume 
there is a set of realizations 
$\HH^0\sb \HH$ of $\P$-measure one 
such that the quenched infinite volume Gibbs measure 
$\mu[\eta]$ is a weak limit (2.2) 
of the quenched finite volume measures (2.1) for all 
$\eta\in \HH^0$. 
Then, a version of the infinite 
volume conditional expectation of the 
corresponding joint measure $K(d\s,d\eta)=\P(d\eta)\mu[\eta](d\s)$
is given by the formula
$$
\eqalign{
&K\left[\x_{\L}\bigl|\x_{\Z^d\ba \L}\right]
= \frac{\mu^{\hbox{\srm ann,}\x_{\del \L}}_{\L}(\x_{\L})}{
\int\mu^{\hbox{\srm ann,}\x_{\del \L}}_{\L}(d\tilde\eta_{\L})
Q_{\L}(\eta_{\L},\tilde\eta_{\L},\eta_{\Z^d\ba\L})
}
}
\tag{3.1}
$$
Here $\mu^{\hbox{\srm ann,}\x_{\del \L}}_{\L}(\x_{\L})$
is the {\bf annealed local specification} given
by (2.7), which can be written in terms of the special 
annealed potential $U^{\srm{ann}}_{A}(\s,\eta)=
\Phi_{A}(\s,\eta)-1_{A=\{x\}}\log \nu(\eta_x)$. 

Further we have put 
$$
\eqalign{
&Q_{\L}(\eta^{1}_{\L},\eta^{2}_{\L},\eta_{\Z^d\ba \L})
:=
\mu[\eta^2_{\L}\eta_{\Z^d\ba \L}]
(e^{-\D H_{\L}(\eta^1_{\L},\eta^2_{\L},\eta_{\del {\L}})})
\cr
}
\tag{3.2}
$$
According to our assumption on the measurability
on $\mu[\eta]$, $Q_{\L}$ depends measurably 
on $\eta_{\Z^d\ba \L}$.
We note the following properties

\item{(i)}
$Q_{\L}(\eta^{1}_{\L},\eta^{2}_{\L},\eta_{\Z^d\ba \L}) 
=\left[
Q_{\L}(\eta^{2}_{\L},\eta^{1}_{\L},\eta_{\Z^d\ba \L})
\right]^{-1}$

\item{(ii)} For any $\D\sp \L$ we have 
$Q_{\D}(\eta^{1}_{\L}\eta_{\D\ba\L},\eta^{2}_{\L}\eta_{\D\ba\L},\eta_{\Z^d\ba \D}) 
=Q_{\L}(\eta^{1}_{\L},\eta^{2}_{\L},\eta_{\Z^d\ba \L})$

\item{(iii)} For any $\eta_{\L}^3$ we have 
$\frac{
Q_{\L}(\eta^{1}_{\L},\eta^{3}_{\L},\eta_{\Z^d\ba \L}) 
}{Q_{\L}(\eta^{2}_{\L},\eta^{3}_{\L},\eta_{\Z^d\ba \L})} 
=Q_{\L}(\eta^{1}_{\L},\eta^{2}_{\L},\eta_{\Z^d\ba \L})$

\noindent whenever $\eta\in \HH^0$.
}

\remark Note that, by our assumption 
on the a.s. convergence of the infinite volume 
Gibbs measures, $Q_{\L}$ can be written in the form
$$
\eqalign{
&Q_{\L}(\eta^{1}_{\L},\eta^{2}_{\L},\eta_{\Z^d\ba \L})
=\lim_{\L_N\uparrow\Z^d}
\mu^{\s^{\hbox{\srm b.c.}}_{\del \L_N}}_{\L_N}
[\eta^2\eta_{\ov{\L_N}}]\left(
e^{-\D H_{\L}(\eta^1_{\L},\eta^2_{\L},\eta_{\del {\L}})}
\right)
=\lim_{\L_N\uparrow\Z^d}\frac{
Z^{\s^{\hbox{\srm b.c.}}_{\del \L_N}}_{\L_N}
[\eta^1_{\L}\eta_{\ov{\L_N}\ba \L}]
}{
Z^{\s^{\hbox{\srm b.c.}}_{\del \L_N}}_{\L_N}
[\eta^2_{\L}\eta_{\ov{\L_N}\ba \L}]
}
}
\tag{3.3}
$$
with the quenched partition function
$$
\eqalign{
&Z^{\s^{\hbox{\srm b.c.}}_{\del \L}}_{\L}
[\eta_{\ov\L}]=
\sum_{\s_{\L}}
e^{-\sum_{A:A\cap \L\neq \em}
\Phi_{A}( \s_{\L}\s^{\hbox{\srm b.c.}}_{\del \L},\eta_{\ov{\L}})
}
}
\tag{3.4}
$$
whenever $\eta\in \HH^0$. Morally, $Q_{\L}$ is thus a fraction between infinite
volume partition functions whose disorder variables 
differ in the volume $\L$.

\remark We note that formulas for the {\it finite
volume conditional expectations} have appeared in [K6] 
[see Lemma 2.1, (2.4) therein]. They seem to look more 
complicated than the infinite volume expression (3.1). 
In that paper we wanted to 
be able to deal also with the more general case 
in which we do not assume $\P$-a.s. convergence of the finite volume 
Gibbs measures,
but only convergence of the finite volume joint measures.  
Then (3.1) is not available. 

\proof
Properties (i),(ii),(iii) are clear from (3.3).

\noindent To get (3.1) we will show  
at first that, for the measure 
$K_{\L_N}^{\s^{\hbox{\srm b.c.}}_{\del\L_N}}(\s_{\L_N},\eta_{\ov{\L_N}})
:= \P(\eta_{\ov{\L_N}})
\mu_{\L_N}^{\s^{\hbox{\srm b.c.}}_{\del\L_N}}[\eta_{\ov{\L_N}}]
(\s_{\L_N})$ on $\O_{\L}\times \HH_{\ov{\L}}$
we have, for finite $\L,\D,\L_{N}$
with $\L\sb \D$ and $\ov{\D}\sb \L_{N}$, the  
formula
$$
\eqalign{
&K_{\L_N}^{\s^{\hbox{\srm b.c.}}_{\del\L_N}}
\left[\x_{\L}\bigl|\x_{\D\ba \L}\right]
= \int 
K_{\L_N}^{\s^{\hbox{\srm b.c.}}_{\del\L_N}}
\left[d\bar\s_{\L_N\ba \D} d\bar\eta_{\ov{\L_N}\ba \D}
\bigl|\x_{\D\ba \L}
\right]
\frac{\mu^{\hbox{\srm ann,}\x_{\del \L\cap \D}\bar \x_{\del \L\ba \D}
}_{\L}(\x_{\L})}{
\int\mu^{\hbox{\srm ann,}\x_{\del \L\cap \D}\bar \x_{\del \L\ba \D}}_{\L}(d\tilde\eta_{\L})
\frac{
Z^{\s^{\hbox{\srm b.c.}}_{\del \L_N}}_{\L_N}
[\eta_{\L}\eta_{\D\ba \L}\bar\eta_{\ov{\L_N}\ba \D}]
}{
Z^{\s^{\hbox{\srm b.c.}}_{\del \L_N}}_{\L_N}
[\tilde\eta_{\L}\eta_{\D\ba \L}\bar\eta_{\ov{\L_N}\ba \D}]
}
}
}
\tag{3.5}
$$
In particular the formula holds true for $\L=\D$. 
Now, (3.4) is just a computation. Indeed, write 
$$
\eqalign{
&K_{\L_N}^{\s^{\hbox{\srm b.c.}}_{\del\L_N}}
\left[\x_{\L}\bigl|\x_{\D\ba \L}\right]\cr
&=\int 
K_{\L_N}^{\s^{\hbox{\srm b.c.}}_{\del\L_N}}
\left[d\bar\s_{\L_N\ba \D} d\bar\eta_{\ov{\L_N}\ba \D}
\bigl|\x_{\D\ba \L}
\right]
K_{\L_N}^{\s^{\hbox{\srm b.c.}}_{\del\L_N}}
\left[\x_{\L}\bigl|\xi_{\D\ba \L}
\bar \s_{\L_N\ba \D}\bar\eta_{\ov{\L_N}\ba \D}\right]
}
\tag{3.6}
$$
and note that the term under the integral on the r.h.s. equals
$$
\eqalign{
&\frac{K_{\L_N}^{\s^{\hbox{\srm b.c.}}_{\del\L_N}}
\left[\x_{\L}\xi_{\D\ba \L}
\bar \s_{\L_N\ba \D}\bar\eta_{\ov{\L_N}\ba \D}\right]}
{\sum_{\tilde \x_{\L}}
K_{\L_N}^{\s^{\hbox{\srm b.c.}}_{\del\L_N}}
\left[\tilde\x_{\L}\xi_{\D\ba \L}
\bar \s_{\L_N\ba \D}\bar\eta_{\ov{\L_N}\ba \D}\right]}\cr
&=
\frac{
\P(\eta_{\L}) 
\mu_{\L_N}^{\s^{\hbox{\srm b.c.}}_{\del\L_N}}[
\eta_{\L}\eta_{\D\ba \L}\bar\eta_{\ov{\L_N}\ba \D}]
( \s_{\L}\s_{\D\ba \L}\bar\s_{\L_N\ba \D})}{
\sum_{\tilde \s_{\L},\tilde\eta_{\L}}
\P(\tilde\eta_{\L}) 
\mu_{\L_N}^{\s^{\hbox{\srm b.c.}}_{\del\L_N}}[
\tilde\eta_{\L}\eta_{\D\ba \L}\bar\eta_{\ov{\L_N}\ba \D}]
(\tilde \s_{\L}\s_{\D\ba \L}\bar\s_{\L_N\ba \D})
}
}
\tag{3.7}
$$
Spelling out the quenched local specifications
in terms of the random potential $\Phi$ this can be 
rewritten in terms of 
the special annealed potential $U^{\srm{ann}}_{A}(\s,\eta)=
\Phi_{A}(\s,\eta)-1_{A=\{x\}}\log \nu(\eta_x)$ as 
$$
\eqalign{
&\frac{
e^{-\sum_{A:A\cap \L\neq \em}U^{\srm{ann}}_{A}(\s_{\L}\s_{\D\ba \L}
\bar\s_{\L_N\ba \D},
\eta_{\L}\eta_{\D\ba \L}\bar\eta_{\ov{\L_N}\ba \D})
}}
{\sum_{\tilde \s_{\L},\tilde\eta_{\L}}
e^{-\sum_{A:A\cap \L\neq \em}U^{\srm{ann}}_{A}(\tilde\s_{\L}\s_{\D\ba \L}
\bar\s_{\L_N\ba \D},
\tilde\eta_{\L}\eta_{\D\ba \L}\bar\eta_{\ov{\L_N}\ba \D})
}
\frac{
Z^{\s^{\hbox{\srm b.c.}}_{\del \L_N}}_{\L_N}
[\eta_{\L}\eta_{\D\ba \L}\bar\eta_{\ov{\L_N}\ba \D}]
}{
Z^{\s^{\hbox{\srm b.c.}}_{\del \L_N}}_{\L_N}
[\tilde\eta_{\L}\eta_{\D\ba \L}\bar\eta_{\ov{\L_N}\ba \D}]
}
}
}
\tag{3.8}
$$
Note that, due to 
cancellations for $\ov{\D}\sb \L_{N} $, the $U$-sums  
do not depend on $\s^{\hbox{\srm b.c.}}$.
Note that, for  $\ov{\L}\sb \D$, (3.8) does not depends 
on $\bar\s_{\L_N\ba \D}$. In this case the outer integral in (3.4)
reduces to an
integration over the disorder variables. 
Note however that this is {\bf not} a product integration!
Finally, normalizing numerator and denominator of (3.8)
by the annealed partition function 
$\sum_{\tilde \s_{\L},\tilde\eta_{\L}}      
e^{-\sum_{A:A\cap \L\neq \em}U^{\srm{ann}}_{A}(\tilde\s_{\L}\s_{\D\ba \L}
\bar\s_{\L_N\ba \D},
\tilde\eta_{\L}\eta_{\D\ba \L}\bar\eta_{\ov{\L_N}\ba \D})
}
$ we get the desired (3.5).

Next we claim that 
$$
\eqalign{
&K\left[\x_{\L}\bigl|\x_{\D\ba \L}\right]
= \int K\left[d\bar\xi_{\Z^d\ba \D}
\bigl|\x_{\D\ba \L}
\right]
\frac{\mu^{\hbox{\srm ann,}\x_{\del \L\cap \D}\bar \x_{\del \L\ba \D}}_{\L}(\x_{\L})}{
\int\mu^{\hbox{\srm ann,}\x_{\del \L\cap \D}\bar \x_{\del \L\ba \D}}_{\L}(d\tilde\eta_{\L})
Q_{\L}(\eta_{\L},\tilde\eta_{\L},\eta_{\D\ba \L}\bar\eta_{\Z^d\ba\D})
}
}
\tag{3.9}
$$
To see this, write down (3.5) explicitly in terms 
of the quenched local specifications and (3.9)
in terms of the infinite volume Gibbs measure.
Note that the dependence on those measures is completely  
local- therefore (3.9) follows by the assumption 
of $\P$-a.s. local convergence of the finite volume 
Gibbs measures. 
But from (3.9) we can conclude now, that what is under 
the integral on the r.h.s. must be the infinite volume
conditional expectation. More precisely, (3.1) follows from the 
following general measure-theoretic

{\bf Fact:} {\it Assume that $\x_{\Z^d}$ 
is a random field with distribution $K$, $\xi_x$ taking
values in a finite set, 
and $\tilde K\left[\x_{\L}\bigl|\x_{\Z^d\ba \L}\right]$
is a Borel probability kernel that satisfies
$$
\eqalign{
&K\left[\x_{\L}\bigl|\x_{\D\ba \L}\right]
= \int K\left[d\bar\xi_{\Z^d\ba \D}
\bigl|\x_{\D\ba \L}
\right]\tilde K\left[\x_{\L}\bigl|\x_{\D\ba \L}\bar\x_{\Z^d\ba \D}\right]
}
\tag{3.10}
$$
for all finite $\D\sp \L$, where $K\left[d\bar\xi_{\Z^d\ba \D}
\bigl|\x_{\D\ba \L}
\right]$ is a version of the conditional expectation. 
Then $\tilde K\left[\x_{\L}\bigl|\x_{\Z^d\ba \L}\right]$
is a version of the infinite volume conditional expectation 
$K\left[\x_{\L}\bigl|\x_{\Z^d\ba \L}\right]$.}

We include a  proof for the convenience of the reader:

$\tilde K\left[\x_{\L}\bigl|\x_{\Z^d\ba \L}\right]$
is assumed to be $\s\left(\xi_{\Z^d\ba \L} \right)$-measurable. 
So, to verify the definition of the conditional 
expectation we have to show that, for all events $C\in \s\left(
\xi_{\Z^d\ba \L} \right)$ and 
$A\in \s\left(\xi_{\Z^d}\right)$ we have that 
$$
\eqalign{
&\int_{C} \left(\int_{A}
\tilde K\left[d\x_{\L}\bigl|\x'_{\Z^d\ba \L}\right]
\otimes \d_{\x'_{\Z^d\ba \L}}(d\x_{\Z^d\ba \L})\right)
K(d\x'_{\Z^d\ba \L})=K(A\cap C)\cr
}
\tag{3.11}
$$
Writing $A$ in the form $A=\sum_{\x_{\L}}\left( 
\{\x_{\L}\}\times A_{\x_{\L}}
\right)$ where $A_{\x_{\L}}\in \s\left(\x_{\Z^d\ba \L} \right)$
we see that this is equivalent to 
$\sum_{\x_{\L}}\int_{C} 
\tilde K\left[\x_{\L}\bigl|\x'_{\Z^d\ba \L}\right]
1_{\x'_{\Z^d\ba \L}\in A_{\x_{\L}}}
K(d\x'_{\Z^d\ba \L})
=\sum_{\x_{\L}}K(\{\x_{\L}\}\times(A_{\x_{\L}}\cap C))
$. 
So, it suffices to show that, for any 
$B\in \s\left(
\xi_{\Z^d\ba \L} \right)$ and any $\x_{\L}$ we have that
$$
\eqalign{
&\int_{B} 
\tilde K\left[\x_{\L}\bigl|\x'_{\Z^d\ba \L}\right]
K(d\x'_{\Z^d\ba \L})
=K(\{\x_{\L}\}\times B)\cr
}
\tag{3.12}
$$
To see this, we apply the standard Dynkin-class argument
to show an equality for all sets of a given $\s$-algebra,  
see e.g. [Co] Theorem 1.6.1 (which 
states that, for any $\cap$-stable set $\FF$ of subsets, the 
smallest $\s$-algebra which contains $\FF$ coincides 
with the smallest Dynkin-class which contains $\FF$). 
First note that the system $\DD$ of sets $B$ in $\s\left(
\xi_{\Z^d\ba \L} \right)$ for which 
this equality holds is a Dynkin class: 
That $\O\in\DD$ follows from (2.10) for $\D=\L$; 
furthermore $\DD$ is stable under formation of 
complements and countable unions of pairwise disjoint 
sets, by the properties of the integral.

Thus we only need to prove (3.12) for
the set of cylinder sets, since they form 
a $\cap$-stable generator of $\s\left(\xi_{\Z^d\ba \L} \right)$.
It suffices to take sets of the form
$B=\{\x, \x_{\D\ba \L}=\x^{(1)}_{\D\ba \L}\}$.
But note that in this case 
$$
\eqalign{
&\int_{B} 
\tilde K\left[\x_{\L}\bigl|\x'_{\Z^d\ba \L}\right]
K(d\x'_{\Z^d\ba \L})
= \int\tilde K\left[\x_{\L}\bigl|\x_{\D\ba \L}\x'_{\Z^d\ba \D}\right]
K(d\x'_{\Z^d\ba \D}|\x_{\D\ba \L})K(\x_{\D\ba\L})\cr
&= K\left[\x_{\L}\bigl|\x_{\D\ba \L}\right]K(\x_{\D\ba\L})
=K(\{\x_{\L}\}\times B)\cr
}
\tag{3.13}
$$
where we have used the hypothesis 
in the second equality.
This concludes the proof of the ``fact'' and 
concludes the proof of the proposition. \endproof

\bigskip
\bigskip

\chap{IV. Construction of Potentials - Proof of the Theorems\hfill}

Starting from the formula of Proposition 3.1 for the infinite 
volume conditional expectations of the joint measure $K$
we will prove Theorems 2.1 and 2.2 at the same time. 
A little later we will prove Theorem 2.4.

As a first consequence of Proposition 3.1 we 
separate the potential for the joint measures we 
are about to construct into an ``annealed part''
and a ``free energy'' part. 
We have  

\lemma{4.1}{\it 
Suppose that $U^{\hbox{\srm ann}}(\xi)$ 
is a potential for the annealed system. 
Then we have that $U(\s,\eta)=U^{\hbox{\srm ann}}(\s,\eta)+
U^{\hbox{\srm fe}}(\eta)$ generates the conditional
expectations for the joint measure $K$ if 
$U^{\hbox{\srm fe}}(\eta)$ is summable for $\P$-a.e $\eta$
and, $\P$-a.s.,  
$$
\eqalign{
&\lim_{\D\uparrow\Z^d}\sum_{A:A\sb \D,A\cap \L\neq \em}
\left(
U^{\hbox{\srm}fe}_A(\eta^1_{\L}\eta_{\Z^d\ba \L})
-U^{\hbox{\srm}fe}_A(\eta^2_{\L}\eta_{\Z^d\ba \L})
\right)
=\log Q_{\L}(\eta^1_{\L},\eta^2_{\L},\eta_{\Z^d\ba \L})
}
\tag{4.1}
$$
}

\proof For finite $\D\sp \L$ we write 
$$
\eqalign{
&\frac{e^{-\sum_{A:A\sb \D,A\cap \L\neq \em}U_A(\xi)}
}{
\sum_{\tilde \x_{\L}}
e^{-\sum_{A:A\sb \D,A\cap \L\neq \em}U_A(\tilde\xi_{\L}\xi_{\Z^d\ba \L})}
}\cr
&=\frac{
e^{-\sum_{A:A\sb \D,A\cap \L\neq \em}U_A^{\hbox{\srm ann}}
(\xi)}
}
{
\sum_{\tilde \x_{\L}}
e^{-\sum_{A:A\sb \D,A\cap \L\neq \em}U_A^{\hbox{\srm ann}}
(\tilde\xi_{\L}\xi_{\Z^d\ba \L})}
e^{-\sum_{A:A\sb \D}
\left(U^{\hbox{\srm}fe}_A(\tilde\eta_{\L}\eta_{\Z^d\ba \L})
-U^{\hbox{\srm}fe}_A(\eta)
\right)}
}\cr
&= \frac{\mu^{\hbox{\srm ann,}\x_{\del \L}}_{\L}(\x_{\L})}{
\int\mu^{\hbox{\srm ann,}\x_{\del \L}}_{\L}(d\tilde\eta_{\L})
e^{-\sum_{A:A\sb \D}
\left(U^{\hbox{\srm}fe}_A(\tilde\eta_{\L}\eta_{\Z^d\ba \L})
-U^{\hbox{\srm}fe}_A(\eta)
\right)}
}
}
\tag{4.2}
$$
Here the first equality is just a resummation 
of sums and the second follows from normalizing by the
annealed partition function. 
Now the claim follows from formula (3.1)
for the infinite volume conditional expectations of $K$
by the limit $\D\uparrow\Z^d$.
\endproof

Thus we are completely reduced to the investigation of the $Q$-part. 
Hence we will define our potentials 
in terms of logarithms of $Q_{\L}$'s. This makes 
life much easier and formulas much more 
transparent than dealing with the full 
conditional probabilities of the joint measures themselves.
The situation is especially nice here, since the $Q$-
part depends only on the disorder variables 
and the marginal of the joint measures we consider
on the disorder variables is just a product measure.

\proofof{Theorem 2.1 and 2.2}
Denote by $\a$ any product-measure on the disorder space. 
Later we will put either $\a=\P$ or $\a=\d_{\hat \eta}$ for 
a fixed realization of the disorder $\hat \eta$,
the first case corresponding to the proof of Theorem 2.1, 
the second case corresponding to the proof of Theorem 2.2. 
For the second case we assume that $\hat \eta$ is 
in the set of realizations for which the convergence (2.2)
holds. From this follows:  
For all realizations which are finite volume perturbations 
of $\hat \eta$ the convergence (2.2) to an infinite volume Gibbs measure 
with the corresponding local specification holds, too.  
(This is seen by splitting off the corresponding terms 
in the Hamiltonian and treating them as a local observable.)
So the l.h.s. of (2.8) is uniquely defined.

We define the `relative energy'
$$
\eqalign{
E^{\a}_{\L}(\eta_{\L})
&:=\int\a(d\tilde\eta)\log Q_{\L}(\eta_{\L},\tilde\eta_{\L},\tilde\eta_{
\Z^d\ba \L})\cr
&= \int\a(d\tilde\eta)\log 
\mu[\tilde\eta]
(e^{-\D H_{\L}(\eta_{\L},\tilde\eta_{\L},\tilde\eta_{\del {\L}})})
\cr
}
\tag{4.3}
$$
and define a potential by the inclusion-exclusion 
principle
$$
\eqalign{
&U^{\hbox{\srm{fe}},\a}_A(\eta)
:= \sum_{\L:\L\sb A}(-1)^{|A\ba \L|}E^{\a}_{\L}(\eta_{\L})
\text{so that }\cr
&E^{\a}_{\L}(\eta_{\L})
=\sum_{A:A\sb \L}U^{\hbox{\srm}fe,\a}_A(\eta)\cr
}
\tag{4.4}
$$
We remark that the application of the inclusion-exclusion principle 
to define a formal potential is a classical thing that 
goes back even before [Koz].
Note that, by choosing $\a=\d_{\hat\eta}$,  
(4.3) becomes an expectation w.r.t. a non-random system 
and thus, 
for a suitable translation-invariant realization $\hat\eta$, 
might even be amenable to explicit computations in certain cases.  
Of course, for $\a=\P$, (4.3) involves the full 
disorder-dependence  of the random Gibbs measure 
and will hardly ever be suitable for explicit computations. 

Note that the family 
of random variables $E^{\a}_{\L}$, indexed 
by finite subsets $\L\sb \Z^d$, is a {\bf martingale w.r.t. 
the product measure $\a$.} This means that, 
for each $\D\sp \L$, 
$$
\eqalign{
&\int \a(d\tilde\eta)E^{\a}_{\D}(\eta_{\L}\tilde\eta_{\D\ba \L})
= E^{\a}_{\L}(\eta_{\L}),\quad
E^{\a}_{\em}
:=\int \a(d\tilde\eta)E^{\a}_{\L}(\tilde\eta_{\L})
= 0
}
\tag{4.5}
$$
Indeed, we have by Proposition 3.1 (iii)
$$
\eqalign{
&\int\a(d\bar\eta)
\int\a(d\tilde\eta)\log Q_{\D}
(\eta_{\L}\bar\eta_{\D\ba\L},\tilde\eta_{\D},\tilde\eta_{Z^d\ba \D})
\cr
&= \int\a(d\bar\eta)
\int\a(d\tilde\eta)
\Bigl(
\log Q_{\D}
(\eta_{\L}\bar\eta_{\D\ba\L},\bar\eta_{\D},\tilde\eta_{Z^d\ba \D})
+\log Q_{\D}
(\bar\eta_{\D},\eta'_{\D},\tilde\eta_{Z^d\ba \D})
+
\log Q_{\D}
(\eta'_{\D},\tilde\eta_{\D},\tilde\eta_{Z^d\ba \D})
\Bigr)\cr
}
\tag{4.6}
$$
for any fixed $\eta'$.
The last two terms cancel, due to Proposition 3.1 (i)
and the first term equals $E^{\a}_{\L}(\eta_{\L})$, 
due to (ii), as desired. Note that this works also 
in the case $\a=\d_{\hat\eta}$ since we assumed weak 
convergence for the point $\hat \eta$!

From this follows easily from the usual play with 
signed sums that, in fact,   
the potential $U^{\hbox{\srm{fe}},\a}$ 
is $\a$-normalized as a potential on the 
disorder space, i.e. $\int\a_x(d\tilde\eta_x) 
U^{\hbox{\srm}fe,\a}_{A}(\eta_{A\ba x}\tilde \eta_x)   
= 0$ whenever $x\in A$. 

Next, to prove that the potential converges, write  
$$
\eqalign{
&\sum_{A:A\sb \D,A\cap \L\neq \em}U^{\hbox{\srm}fe,\a}_A(\eta)
=\sum_{A:A\sb \D}U^{\hbox{\srm}fe,\a}_A(\eta)
-\sum_{A:A\sb\D\ba \L}U^{\a,\hbox{\srm}fe}_A(\eta) \cr
&=E^{\a}_\D(\eta)-E^{\a}_{\D\ba \L}(\eta) \cr
&=\int\a(d\tilde\eta) 
\log\frac{Q_{\D}(\eta_{\D},\tilde\eta_{\D},\tilde\eta_{\Z^d\ba \D})}{
Q_{\D}(\tilde\eta_{\L}\eta_{\D\ba\L},\tilde\eta_{\D},
\tilde\eta_{\Z^d\ba \D})
}\cr
&=\int\a(d\tilde\eta)\log Q_{\L}(\eta_{\L},\tilde\eta_{\L},\eta_{\D\ba\L}
\tilde\eta_{\Z^d\ba \D})\cr
}
\tag{4.7}
$$
The second equality is (4.4) and for the next
two equalities we have used properties (ii) and (iii)
for $Q$.
The important point that exploits the nature 
of $\a$ being a product measure is 
the convergence statement 
$$
\eqalign{
&\lim_{\D\uparrow\Z^d}\int\a(d\tilde\eta)
\log Q_{\L}(\eta^1_{\L},\eta^2_{\L},
\eta_{\D\ba \L}\tilde\eta_{\Z^d\ba \D})
= \log Q_{\L}(\eta^1_{\L},\eta^2_{\L},
\eta_{\Z^d\ba \L}) \text{for}
\a\hbox{-a.e. }\eta 
}
\tag{4.8}
$$
This follows by the martingale convergence theorem, since,
for any fixed finite $\L\sb \Z^d$ and fixed $\eta_{\L}^1,
\eta^2_{\L}$
 the expression under the limit on the l.h.s 
indexed by finite subsets 
$\D\sb \Z^d$ s.t. $\D\sp \L$, is a martingale w.r.t the 
distribution given by $\a$.

\noindent {\bf Theorem 2.1:} We put $\a=\P$. Then we see from (4.7)
and (4.8) that the potential converges with $\D\uparrow\Z^d$
for $\P$-a.e. $\eta$. 
Since $\P$ is the marginal of $K$ on the disorder-space, 
this is exactly what we want. 

\noindent {\bf Theorem 2.2:} We put $\a=\d_{\hat \eta}$
where $\hat \eta$ is the assumed direction of continuity. 
In this case the r.h.s. of (4.7) is just 
$Q_{\L}(\eta_{\L},\hat\eta_{\L},\eta_{\D\ba\L}
\hat\eta_{\Z^d\ba \D})$. Using property (iii) for $Q_{\L}$
we may rewrite this as a telescoping sum 
$\sum_{x\in \L}
Q_{\L}(\eta_{\L_{\leq x}},\eta_{\L_{< x}},\eta_{\D\ba\L}
\hat\eta_{\Z^d\ba \D})$. Here 
we have put the lexicographic order on $\Z^d$ and written
$\L_{\leq x}=\{z\in \L; z\leq x\}$ (and the analogous 
notation for ``$<$''). Thus we see that (2.7)
really implies convergence of the potential with $\D\uparrow\Z^d$.

Next we prove that the potential generates the infinite 
volume conditional expectations of the joint measure $K$.
We must verify hypothesis (4.1) of Lemma 4.1. We have 
$$
\eqalign{
&
\sum_{A:A\sb \D}
\left(
U^{\hbox{\srm}fe}_A(\eta^1_{\L}\eta_{\Z^d\ba \L})
-U^{\hbox{\srm}fe}_A(\eta^2_{\L}\eta_{\Z^d\ba \L})
\right)\cr
&=E^{\a}_\D(\eta^1_{\L}\eta_{\D\ba \L})
-E^{\a}_{\D}(\eta^2_{\L}\eta_{\D\ba \L}) \cr
&=\int\a(d\tilde\eta_{\Z^d})\log\frac{
Q_{\D}(\eta^1_{\L}\eta_{\D\ba\L},\tilde\eta_{\D},\tilde\eta_{\Z^d\ba\D})}
{Q_{\D}(\eta^2_{\L}\eta_{\D\ba\L},\tilde\eta_{\D},\tilde\eta_{\Z^d\ba \D})}
\cr
&=\int\a(d\tilde\eta_{\Z^d})\log
Q_{\L}(\eta^1_{\L},\eta^2_{\L},\eta_{\D\ba\L}\tilde\eta_{\Z^d\ba \D})
\cr
}
\tag{4.9}
$$
But, recalling (4.8),  the proof of (4.1)
is the same as that of the convergence of the potential, 
in the respective cases of Theorem 2.1 and Theorem 2.2. 
This concludes the proof of Theorems 2.1 and 2.2. 
The convergence statement of Corollary 2 follows from (4.7)
by integration over $\eta$ w.r.t. $\P$. In fact, we see that 
$\sum_{A:A\cap \L\neq \em}
\int\P(d\tilde\eta)V^{\hbox{\srm{fe}}}_{\mu; A}(\tilde\eta)$
equals the finite quantity 
$ \int \P(d\eta)\log Q_{\L}(\tilde\eta_{\L},\hat\eta_{\L},
\tilde\eta_{\Z^d\ba \L})$. 
Finally we also note that, assuming continuity of $Q$
everywhere, we have even pointwise convergence of (4.8)
for both choices of $\a$. This proves the 
first convergence statement after (2.9). 
\endproof

\bigskip

\line{\bf A general remark about resummed potentials:\hfill}

The potentials used in the proofs of Theorem 2.3 and 
Theorem 2.4 are obtained by resumming the 
supports of the $\a$-normalized 
potential $U_{A}^{\a,\hbox{\srm fe}}(\eta)$. 
The general construction is the following:
Denote by $P$ the set of finite subsets of $\Z^d$
and let $P=\bigcup_{a} P_a$ be a disjoint union 
s.t. (i) $C_a:= \bigcup_{A:A\sb P_a} A$ is finite for every $a$,
and (ii) there exists a net of finite sets $\D_{\b}\sb \Z^d$ 
s.t. $\lim_{\b}\D_{\b} =\Z^d$ and: 
for all finite $\L$, we have that, for sufficiently large
$\D_{\b}$, for all $A\sb \D_\b$ s.t. $A\cap \L\neq \em$
there exists an $a$ with $C_a\sb \D_\b$ 
s.t. $A\in P_a$. Then $U_{C}^{\a,\hbox{\srm fe, gr}}(\eta)$, defined by  
$$
\eqalign{
&U_{C_a}^{\a,\hbox{\srm fe, gr}}(\eta)
:=\sum_{A:A\sb P_a } U_{C}^{\a,\hbox{\srm fe}}(\eta), \quad 
U_{C}^{\a,\hbox{\srm gr}}(\eta):=0 \hbox{ if }
C\neq C_a \hbox{ for all }a
}
\tag{4.10}
$$
is called the 
{\bf  resummed potential } corresponding to the 
given decomposition of $P$. The reason for 
the complicated looking requirement (ii) is that one has 

\lemma{4.2}{\it 
Suppose that $U_{C}^{\a,\hbox{\srm fe, gr}}(\eta)$ 
is a resummed potential obtained from the 
$\a$-normalized free energy potential $U_{C}^{\a,\hbox{\srm fe}}(\eta)$
that converges {\bf absolutely} for 
$\P$-a.e. $\eta$.
Then $U(\s,\eta)=U^{\hbox{\srm ann}}(\s,\eta)+
U^{\a,\hbox{\srm fe, gr}}(\eta)
$ generates the conditional
expectations for the joint measure $K$ (for any annealed potential),
if the $\a$-normalized potential does. 
}

\proof 
For any fixed $\L$ we have that, for any sufficiently large $\D_{\b}$,
$$
\eqalign{
&\sum_{C:C\sb \D_{\b},C\cap \L\neq \em}
\left(
U_{C}^{\a,\hbox{\srm fe, gr}}(\eta^1_{\L}\eta_{\Z^d\ba \L})
-U_{C}^{\a,\hbox{\srm fe, gr}}(\eta^2_{\L}\eta_{\Z^d\ba \L})
\right)\cr
&=
\sum_{A:A\sb \D_{\b},A\cap \L\neq \em}
\left(
U_{A}^{\a,\hbox{\srm fe}}(\eta^1_{\L}\eta_{\Z^d\ba \L})
-U_{A}^{\a,\hbox{\srm fe}}(\eta^2_{\L}\eta_{\Z^d\ba \L})
\right)\cr
}
\tag{4.11}
$$
This is clear, since, for every term in the right sum
there is precisely one term in the left sum 
containing its contribution, due to property (ii). 
Conversely, those contributions 
on the l.h.s. coming from $A$'s that don't intersect $\L$
cancel because the field configurations agree outside of $\L$. 
Thus, the l.h.s. converges to the r.h.s. of (4.1) along 
the net $\D_{\b}$. By the hypothesis of absolute convergence
this implies convergence for any sequence $\D\uparrow\infty$,
which proves the claim, by Lemma 4.1.\endproof 

The resummations used in the proofs of Theorem 2.3 and 2.4 
were invented already by [Koz] and used in various 
publications since then. There are of the following general form. 
Take $\leq$ any total order of the lattice 
points in $\Z^d$. Let, for any lattice point $x\in \Z^d$,
an increasing  sequence of finite 
subsets $A_{x,m}\sb \{y:y\geq x\}$, 
$m=1,2,\dots$ be given s.t. $\bigcup _{m } A_{x,m}=\{y:y\geq x\}$. 
Put $A_{x,m=0}=\em$ and define 
$P_{x,m}:=\{A:x\in A\sb A_{x,m},
A\cap ( A_{x,m}\ba A_{x,m-1})\neq \em \}$.
The second condition for the sum is empty  
for $m=1$. Then $\bigcup_{x,m}P_{x,m}=P$ 
is a disjoint union and condition (i) is satisfied. Indeed, 
to see (ii), take the family
$\D_{\un m}=\bigcup_{x\in \Z^d} A_{x,m_x} $
where $\un m=(m_x)_{x\in \Z^d}$ is an integer 
vector s.t. only finitely many of the $m_x$'s are nonzero.

\proofof{Theorem 2.3} 
By Lemma 4.2 it suffices  to show a.s. summability 
of a certain resummed potential.  
The proof of this statement essentially relies 
on an $L^1$-statement corresponding to 
the convergence result (4.8). 
In order to explain why this ensures 
the existence of an a.s. summable potential, however, 
we have to write down explicit formulas.
Let $x\mapsto \#(x)$ denote a one-to-one map from 
$\Z^d$ to the integers $\{1,2,\dots\}$. (The reader 
may think of some spiraling order.) 
Then the $L^1$-martingale convergence theorem gives us that
$$
\eqalign{
&
\int\P(d\eta)\Bigl|
\int\P(d\tilde\eta)
\log Q_{x}(\eta_{x},\tilde\eta_x, 
\tilde \eta_{\{y:1\leq \#(y)<\#(x)\}}   
\eta_{\{y:\#(x)< \#(y)\leq r\}}
\tilde \eta_{\{y:\#(y)> r\}})    \cr
&\quad -\int\P(d\tilde\eta)
\log Q_{x}(\eta_{x},\tilde\eta_x, 
\tilde \eta_{\{y:1\leq \#(y)<\#(x)\}}   
\eta_{\{y:\#(y)>\#(x)\}})
\Bigr|=: \e_x(r)\downarrow 0
\cr
}
\tag{4.12}
$$
with $r\uparrow\infty$, for any fixed $x$. 
This is clear, since the first line 
of the expression under the modulus  is a martingale
w.r.t. to the parameter r, for any fixed $x$ and fixed $\eta_x$.

Take some subsequence $r(n)$ of the integers, to be defined below.  
For $x\geq 1$, $m\geq 1$ 
define $A_{x,m}:=\{z\in \Z^d, \#(x)\leq \#(z) \leq r(x+m)\}$, 
put also $A_{x,m=0}=\em$. 
Starting from general $\a$, let us 
define the resummed potential by the formula
corresponding to (4.10), i.e. 
$$
\eqalign{
&U_{A_{x,m}}^{\a,\hbox{\srm fe, abs}}(\eta)
:=\sum_{{A:x\in A\sb A_{x,m}}\atop{
A\cap ( A_{x,m}\ba A_{x,m-1})\neq \em}}U_{A}^{\a,\hbox{\srm fe}}(\eta),
\quad U_{C}^{\a,\hbox{\srm fe, abs}}(\eta)=0 \hbox{  otherwise}
}
\tag{4.13}
$$
for all $x\in \Z^d$ and $m\geq 1$. Then we have for $m\geq 2$
$$
\eqalign{
&U_{A_{x,m}}^{\a,\hbox{\srm fe, abs}}(\eta)
= E^{\a}_{A_{x,m} }(\eta)
- E^{\a}_{A_{x,m-1} }(\eta)
-E^{\a}_{A_{x,m}\ba x }(\eta)
+ E^{\a}_{A_{x,m-1}\ba x }(\eta)\cr
&= 
\int\a(d\tilde\eta_{\Z^d})\log\frac{
Q_{A_{x,m}}(\eta_{A_{x,m}},
\tilde\eta_{A_{x,m}},\tilde\eta_{\Z^d\ba A_{x,m}})
Q_{A_{x,m}}(\eta_{A_{x,m-1}\ba x }
\tilde\eta_{A_{x,m}\ba (A_{x,m-1}\ba x )},
\tilde\eta_{A_{x,m}}
,\tilde\eta_{\Z^d\ba A_{x,m}})
}
{Q_{A_{x,m}}(\eta_{A_{x,m-1}} \tilde\eta_{A_{x,m}\ba A_{x,m-1} }  ,
\tilde\eta_{A_{x,m}},\tilde\eta_{\Z^d\ba A_{x,m}})
Q_{A_{x,m}}(\eta_{A_{x,m}\ba x}\tilde \eta_x,
\tilde\eta_{A_{x,m}},\tilde\eta_{\Z^d\ba A_{x,m}})
}
}
\tag{4.14}
$$
In the first line we have used the expression of the 
relative energies in terms of the potential. 
In the last line we have used the definition of the 
relative energies and property (iii) for $Q$. 
Again, by (iii), this can be rewritten as 
$$
\eqalign{
&
U_{A_{x,m}}^{\a,\hbox{\srm fe, abs}}(\eta)=\int\a(d\tilde\eta)
\log\frac{
Q_{x}(\eta_{x},\tilde \eta_x, \eta_{A_{x,m}\ba x }
\tilde\eta_{\Z^d\ba  A_{x,m}})}
{
Q_{x}(\eta_{x},\tilde \eta_x, \eta_{A_{x,m-1}\ba x }
\tilde\eta_{\Z^d\ba  A_{x,m-1}})}
}
\tag{4.15}
$$
The previous formula was true 
for any resummed potential starting from the $\a$-normalized
free energy potential. 
Let us switch to $\a=\P$ and drop the subscript $\a$. 
Now we have from the convergence property (4.12)
our main estimate: 
$$
\eqalign{
&\int\P(d\tilde\eta)\left |
U_{A_{x,m}}^{\hbox{\srm fe, abs}}(\tilde\eta)\right|
\leq 2 \e_{x}(r(x+m-1))
\cr
}
\tag{4.16}
$$
Similar to (4.14), (4.15) we have for $m=1$
$$
\eqalign{
&U_{A_{x,1}}^{\a,\hbox{\srm fe, abs}}(\eta)
= E^{\a}_{A_{x,1} }(\eta)
-E^{\a}_{A_{x,1}\ba x }(\eta)\cr
&=\int\a(d\tilde\eta)\log\frac{
Q_{A_{x,1}}(\eta_{A_{x,1}},
\tilde\eta_{A_{x,1}},\tilde\eta_{\Z^d\ba A_{x,1}})}
{Q_{A_{x,1}}(\eta_{A_{x,1}\ba x}\tilde \eta_x,
\tilde\eta_{A_{x,1}},\tilde\eta_{\Z^d\ba A_{x,1}})
}\cr
&= \int\a(d\tilde\eta)\log
Q_{x}(\eta_{x},
\tilde\eta_{x}, \eta_{A_{x,1}\ba x}\tilde\eta_{\Z^d\ba A_{x,1}})
}
\tag{4.17}
$$
This is uniformly bounded in modulus by some constant $\Const_{1}$.
From the last two estimates one concludes that 
$$
\eqalign{
&\sum_{C:C\ni x}
\int\P(d\tilde\eta)\left |
U_{C}^{\hbox{\srm fe, abs}}(\tilde\eta)\right|
\leq \sum_{y:\#(y)\leq \#(x)}
\sum_{m=1}^{\infty}
\int\P(d\tilde\eta)
\left| U_{A_{x,m}}^{\hbox{\srm fe, abs}}(\tilde\eta)\right|\cr
&\leq \Const_1 |\{y,\#(y)\leq \#(x)\}|
+ 2 \sum_{y:\#(y)\leq \#(x)}\sum_{m=2}^\infty
\e_y(r(\#(y)+m-1))
\cr
}
\tag{4.18}
$$
But, it is a simple matter to convince oneself 
that it is possible to choose a subsequence
$r(m)$ of the integers s.t. the $m$-sum 
is finite for all $y$. 
(In fact, from $\e_x(r)\downarrow 0$ 
one can find a subsequence $r(n)$
s.t. even $\sum_{n=1}^\infty \e_{y}(r(n))$.)
This completes the definition of the potential 
and proves $\P$-integrability and thus, in particular, 
$\P$-a.s. summability.\endproof  

The readers may check for themselves 
that one may rerun the proof for both choices 
of $\a$ under the hypothesis of continuity 
of $Q$ everywhere. This proves the 
strengthened version of Theorem 2.3 promised 
after (2.9).  
One may however {\bf not } rerun the proof
for $\a=\d_{\hat \eta}$ without further 
assumptions other than the continuity of $Q_x$ in the 
direction $\hat \eta$ with the hope to obtain 
an absolutely summable potential. 
This is because the speed of convergence 
of the analogue of (4.12) (obtained by replacing 
$\P$ by $\d_{\hat \eta}$)
may be nonuniform in $\eta$ in this case.

\bigskip

\proofof{Theorem 2.4}
 
This time, denote $A_{x,m}:=\{z\in \Z^d;z\geq  x, |z-x|\leq m \}$  
and define the potential by the same formula (4.13), with 
the new $A$'s. Then (4.15) and (4.17) stay true.
(4.17) is uniformly bounded.  
The potential can be rewritten in terms of
correlations. Introduce 
$Q_{x,m,\leq y}:=L_{x,m-1}\cup \{z\in L_{x,m}\ba L_{x,m-1}; z\leq y \}$.
Then, for $m\geq 2$ we have  
$$
\eqalign{
&U^{\hbox{\srm{fe,abs,inv}}}_{L_{x,m}}(\eta)
=\sum_{y\in L_{x,m}\ba L_{x,m-1}}\left(
E^{\a}(\eta^{Q_{x,m,\leq y} })
- E^{\a}(\eta^{Q_{x,m,<y} })
-E^{\a}(\eta^{Q_{x,m,\leq y}\ba x })
+ E^{\a}(\eta^{Q_{x,m,<y}\ba x })
\right) 
\cr
}
\tag{4.19}
$$
The term in brackets can be expressed as 
$$
\eqalign{
& -\int \a(d\tilde\eta)\log\frac
{\mu[\eta^{Q_{x,m,<y}\ba x }]
\left(
e^{-\D H_{\{x,y\}}(\eta_{\{x,y\}},\tilde\eta_{\{x,y\}},\eta^{Q_{x,m,<y}\ba x }
\bigl|_{\del {\{x,y\}}})
}
\right)}
{
\mu[\eta^{Q_{x,m,<y}\ba x }]
\left(
e^{-\D H_{x}(\eta_{x},\tilde\eta_{x},\eta^{Q_{x,m,<y}\ba x }
\bigl|_{\del {x}})}
\right)
\mu[\eta^{Q_{x,m,<y}\ba x }]
\left(
e^{-\D H_{y}(\eta_{y},\tilde\eta_{y},\eta^{Q_{x,m,<y}\ba x }
\bigl|_{\del {y}})}
\right)
}
}
\tag{4.20}
$$
where we have used the notation $\eta^A:=(\eta_A\tilde\eta_{\Z^d\ba A})$.
Note that this gives a $\tilde\eta$-dependence for the $\a$-integral.
So we get that $\eta$-expectation
of  the modulus of the l.h.s. is bounded from above by 
$$
\eqalign{
&\int\a(d\eta)\left|E^{\a}(\eta^{Q_{x,m,\leq y} })
- E^{\a}(\eta^{Q_{x,m,<y} })
-E^{\a}(\eta^{Q_{x,m,\leq y}\ba x })
+ E^{\a}(\eta^{Q_{x,m,<y}\ba x })\right|
\cr
&\leq \Const \int \a(d \eta)\Biggl|
\int \a(d \tilde\eta)\mu[\eta^{Q_{x,m,<y}\ba x }]
\left(e^{-\D H_{\{x,y\}}(\eta_{\{x,y\}},\tilde\eta_{\{x,y\}},\eta^{Q_{x,m,<y}\ba x }
\bigl|_{\del {\{x,y\}}})}\right)\cr
&\quad-\int \a(d \tilde\eta)\mu[\eta^{Q_{x,m,<y}\ba x }]
\left(
e^{-\D H_{x}(\eta_{x},\tilde\eta_{x},\eta^{Q_{x,m,<y}\ba x }
\bigl|_{\del {x}})}
\right)
\mu[\eta^{Q_{x,m,<y}\ba x }]
\left(
e^{-\D H_{y}(\eta_{y},\tilde\eta_{y},\eta^{Q_{x,m,<y}\ba x }
\bigl|_{\del {y}})}
\right)\Biggr|\cr
&
}
\tag{4.21}
$$
where, as always, we have used that $\D H_x$ is uniformly bounded
to drop the logarithm. 
Let us now switch to the case $\a=\P$. 
We use the inequality $|\int f|\leq \int|f|$  
for the $\tilde \eta$-integration to see  that 
the r.h.s.  is bounded from above by 
$\Const \int\P(d\tilde \eta)
\left| c_{x,y}(\eta_x,\eta_y,\tilde \eta)\right|$,
the latter quantity being defined in (2.10). 
Recalling $\bar c(m):=\sup_{{x,y:|x-y|=m}\atop{
\eta_x,\eta_y\in \HH_0}} 
\int\P(d\tilde \eta)
\left| c_{x,y}(\eta_x,\eta_y,\tilde \eta)\right|$ 
we have from this and (4.21) that
$$
\eqalign{
&\int\P(d\eta)\left|U^{\hbox{\srm{fe,abs,inv}}}_{L_{x,m}}(\eta) \right|
\leq\Const\left|L_{x,m}\ba L_{x,m-1}\right|\bar c(m)\leq 
\Const' m^{d-1}\bar c(m)
}
\tag{4.22}
$$
But this gives
$$
\eqalign{
&\sum_{A:A\ni x_0} w(A)
\int\P(d\tilde\eta)
\left |
U^{\hbox{\srm{fe,abs,inv}}}_{A}(\tilde \eta)\right|\cr
&\leq \sum_{m=1}^{\infty} 
\sum_{y:|y-x_0|\leq m}
w(A_{y,m})
\int\P(d\tilde\eta)
\left |
U^{\hbox{\srm{fe,abs,inv}}}_{A_{y,m}}(\tilde \eta)\right|
\leq \Const_1 +\Const_2
\sum_{m=2}^{\infty} m^{2d-1} w(A_{0,m}) \bar c(m)
\cr
}
\tag{4.23}
$$
which finishes the proof. 
\endproof

We remark that the trick to relate some formal potential 
to expectations of certain observables by a telescoping 
[as in (4.19), (4.20)]
was used in various papers before.  
Observe e.g. the analogy to the recent [MRSM] where a.s. 
strongly decaying potentials 
for renormalized measures of low temperature spin systems
were constructed.

\vfill\eject

\chap{V. Examples}

The results of Theorems 2.1 and 2.3 are 
general existence results that always apply.  
Let us however also see what the more specific 
assumptions needed for the convergence of the 
vacuum potential and the strengthenings of 
Theorems 2.1,2.3 given after (2.9) and in Theorem 2.4 
mean in the examples of the (i) random field Ising model, 
(ii) Ising models with random couplings, and the 
(iii) diluted Ising ferromagnet.  
These examples were discussed already in [K6]
w.r.t the question of almost Gibbsianness.   

\noindent{\thbf (i) The Random-Field Ising Model:} 
The single spin space for the variables $\s_x$ is $\O_0=\{-1,1\}$.
The disorder variables are given by the random fields $\eta_x$ 
that are i.i.d. with single-site distribution $\nu$
that is supported on a finite set $\HH_0$ and assumed 
to be symmetric. 
The disordered potential $\Phi(\s,\eta)$ is 
given by $\Phi_{\{x,y\}}(\s,\eta)=-J\s_x\s_y$
for nearest neighbors $x,y\in \Z^d$, 
$\Phi_{\{x\}}(\s,\eta)=-h\eta_x\s_x$,  and $\Phi_A=0$ else. 
Note that 
$e^{-\D H_{x}(\s_{x},\eta_{x}^1,\eta_{x}^2)}= e^{h(\eta^1_x-\eta_x^2)\s_x}
=e^{h(\eta_x^2-\eta_x^1)}+2\sinh h(\eta_x^1-\eta_x^2)\,\, 1_{\s_x=1}
$. 
Then, treating this exponential as an observable and 
using the `finite volume perturbation formula' as in [K6] 
we see the following. 
Condition (2.8) (giving the convergence 
of the vacuum potential) holds if and only if 
$$
\eqalign{
&\lim_{\L\uparrow\Z^d}\mu[\eta_{\L}\hat\eta_{\Z^d\ba \L}]
(\tilde \s_x=1)
=\mu[\eta](\tilde \s_x=1)
\cr
}
\tag{5.1}
$$
for $\eta_x$, for all $x$, for $\P$-a.e. $\eta$.
(Here, as always, we used the notation 
that spins that are integrated are decorated with tildes.)
This is true for any measurable infinite volume 
Gibbs measure  $\mu[\eta]$ which is obtained as 
a weak limit with a non-random boundary condition.
We note that whether (5.1) holds is independent 
of $\eta_x$.
Similarly, condition (2.9) (giving continuity 
of the conditional expectations) holds, whenever 
$$
\eqalign{
&\lim_{\L\uparrow\Z^d}\sup_{\hat\eta}\left|
\mu[\eta_{\L}\hat\eta_{\Z^d\ba \L}]
(\tilde \s_x=1)
- \mu[\eta]
(\tilde \s_x=1)\right|
=0
}
\tag{5.2}
$$
From this we have 

\noindent{\bf Corollary to Theorem 2.2: }{\it 
For any choice of the parameters 
of the model, the joint measure corresponding 
to the ferromagnetic plus-state has a convergent vacuum 
potential with vacuum $(\eta^+,\s)$. Here $\eta^+$
is the configuration taking the maximum of the possible 
values of the magnetic field for all sites $x$ and $\s$
is an arbitrary spin-configuration.}

\noindent{\bf Corollary to Theorems 2.1,2.3: }{\it 
Suppose that $\lim_{\L\uparrow\Z^d}\mu^+_{\L}[\eta_{\L}](\tilde\s_x=1)
=\lim_{\L\uparrow\Z^d}\mu^-_{\L}[\eta_{\L}](\tilde\s_x=1)$ 
for all choices of the magnetic fields $\eta\in \HH$.  
Here the expressions under the limit refer to 
the finite volume Gibbs-measures with $+$ (resp. $-$) 
boundary condition. 

Then the corresponding (unique) joint measure is Gibbs
and the potentials of Theorems 2.1 and 2.2 are  
both convergent everywhere. 
There is also a potential of the form 
announced in Theorem 2.3 that is 
absolutely convergent everywhere.}

\noindent{\bf Proof of Corollaries: } 
It is known that the limit $\mu^+[\eta]=
\lim_{\L\uparrow\Z^d}\mu_{\L}^+[\eta_{\L}]$
exists for any
choice of the parameters and any configuration
of the quenched random fields $\eta_x$, due to 
monotonicity reasons. To prove the first Corollary
we show that (5.1)
holds for $\mu^+$ and $\hat\eta=\eta^+$ and {\bf any} $\eta$.
To see this use the fact that the function 
$(\eta,\s^{\hbox{\srm bc}})\mapsto
\mu^{\s^{\hbox{\srm bc}}}_{\L}
[\eta_{\L}]\left(\tilde \s_x=1\right)$
is monotone (w.r.t. the partial order of its arguments obtained by
site-wise comparison.) 
From this we have 
$$
\eqalign{
&\mu^+[\eta](\tilde \s_x=1)=
\limsup_{\L\uparrow\Z^d}\mu_{\L}^+[\eta_{\L}](\tilde \s_x=1)
\geq \limsup_{\L\uparrow\Z^d}\mu[\eta_{\L}\eta^+_{\Z^d\ba \L}](\tilde \s_x=1)
\cr
}
\tag{5.3}
$$
for any $\eta$ where inequality under the limsup follows
from the DLR-equation and the monotonicity. 
Additionally we have the converse estimate that follows from  
$$
\eqalign{
&
\mu^+[\eta](\tilde \s_x=1)
= 
\lim_{\L_2\uparrow\Z^d}\mu_{\L_2}^+[\eta_{\L_2}](\tilde \s_x=1)
\leq \lim_{\L_2\uparrow\Z^d}\mu_{\L_2}^+[ \eta_{\L}\eta^+_{\L_2\ba \L}]
(\tilde \s_x=1)
= \mu[\eta_{\L}\eta^+_{\Z^d\ba \L}](\tilde \s_x=1)
\cr
}
\tag{5.4}
$$
by taking the $\liminf$ over $\L$. 
This proves the claim. The other Corollary follows from 
the remark after (2.9) and the fact that (5.2)
follows from the hypothesis by $
\mu_{\L}^-[\eta_{\L}](\tilde \s_x=1)\leq 
\mu[\eta_{\L}\hat\eta_{\Z^d\ba \L}]
(\tilde \s_x=1)
\leq \mu_{\L}^+[\eta_{\L}](\tilde \s_x=1)$
for any $\hat\eta$.\endproof  

\bigskip

Next we discuss the hypothesis of Theorem 2.4 giving 
decay of a translation invariant potential. 
Again, using the special form of the single-site perturbation 
of the Hamiltonian, 
it is not difficult to see that we have 
$$
\eqalign{
&\bar c(m)\leq 
\Const 
\sup_{{x,y:|x-y|=m}}
\int\P(d\tilde \eta)
\left|\mu[\tilde \eta](\tilde\s_x \tilde\s_y)
-\mu[\tilde \eta](\tilde\s_x)
\mu[\tilde \eta](\tilde\s_y)
\right|
\cr
}
\tag{5.5}
$$
for $m\geq 1$. (Here the sup over the 
possible different choices of $\eta_x$ and $\eta_y$
was absorbed in the constant. To see this 
we used the `finite volume perturbation formula'
as in [K6] Chapter III.1)  

Now, let us assume that we are in the interesting region 
of the parameter space where existence of 
ferromagnetic order is proved. I.e, let us assume 
that we are in dimensions $d\geq 3$  and we have 
small disorder and large temperature, i.e. $J>0$ sufficiently large 
and $h/J$ is sufficiently small. Then, 
a refined analysis of the renormalization group 
proof of Bricmont and Kupiainen should lead to the fact that 
(5.5) decays faster than any 
power with $m\uparrow\infty$ for the plus-state $\mu^+[\eta]$. 
[Unfortunately this does not follow directly from 
the (related) statement (2.6) given under [BK] Theorem (2.1)
which asserts that the quenched correlation 
under the $\P$-integral decays like 
$\Const(\tilde\eta)e^{- \const |x-y|}$, 
since $\Const(\tilde\eta)$ is unbounded.]  
This has to be contrasted with the 
fact that in this region the system was already 
proved to be {\bf not} almost Gibbsian in [K6]. 
(The set of ``bad configurations'' of $\eta$ 
even has full measure. The reason for 
this is that the magnetization $\mu^+[\eta](\tilde \s_x)$ 
can be made to jump for typical $\eta$ 
by varying the signs of the 
field $\eta$ in a large annulus arbitrarily far 
away from $x$. So, (5.2) does certainly not hold.)

In the opposite "high temperature" 
case where the coupling $J$ is sufficiently small, 
one gets exponential decay $\bar c(m)\leq   
\Const e^{-\const |x-y|}$. In fact, stronger than that, 
one has an exponential bound 
on the random correlations in (5.5), uniformly 
in all realizations of the field. 
For small $J$ this can be seen by a 
standard expansion of the nonrandom interaction 
term  $e^{J 1_{\s_x=\s_y}}= e^{J 1_{\s_x=\s_y}}-1+1$. Indeed, 
summation over the spins w.r.t. the independent measures 
$\nu(d\s_x)e^{h\eta_x\s_x}$ then produces an $\eta$-dependent 
polymer model that has exponential decay of correlations, 
uniformly in $\eta$.
Of course, exponential decay of quenched correlations, uniformly 
in the realization of the fields, always holds in one 
dimension. This can be seen (e.g.) 
by disagreement percolation arguments. 
By the remark after Theorem 2.4 this implies that 
the joint measure is Gibbsian with an interaction 
potential that is superpolynomially 
decaying everywhere.  


\noindent{\thbf (ii) Ising Models with Random Nearest Neighbor Couplings: 
Random Bond, EA-Spinglass: } 
The single spin space is again $\O_0=\{-1,1\}$.
Denote by $\EE:=\{(1,0,0,\dots,0),$
$(0,1,0,\dots,0),\dots,(0,0,\dots,1)\}$ the set 
of nearest neighbor vectors pointing in `positive directions'. 
The disorder variables (random couplings) 
$J_{x,e}$ take finitely many values, independently
over the `bonds' $x,e$. We put $\eta_x=(J_{x,e})_{e\in \EE}$.  
The {\bf joint spin} at the site $x$
is then $\xi_x=(\s_x,\eta_x)=(\s_x,(J_{x,e})_{e\in \EE})$.
The disordered potential $\Phi(\s,\eta)$ is 
given by $\Phi_{\{x,y\}}(\s,\eta)=-J_{x,e}\s_x\s_y$ 
if $y=x+e$ for some $e\in \EE$, and $\Phi_A=0$ else. 
Specific distributions of interest are 
a) $J_{x,e}$ takes values strictly bigger than zero 
(random bond ferromagnet); 
b) $J_{x,e}$ is symmetrically distributed 
(EA-spinglass).

Now, the crucial observable is the correlation 
between nearest neighbors. We use
the special form of the single site perturbation 
of the Hamiltonian w.r.t. $\eta_x$ and similar 
arguments as for the random field Ising model
(see [K6] chapter III.3). In this way we see that: (2.8) holds if 
$$
\eqalign{
&\lim_{\L^*\uparrow(\Z^d)^*}
\mu_{\infty}[J_{\L^*}
\hat J_{(\Z^d)^*\ba\L^*}](\tilde \s_x
\tilde\s_y)
= \mu_{\infty}[J_{(\Z^d)^*}](\tilde \s_x
\tilde\s_y)
}
\tag{5.6}
$$
for any nearest neighbor pair $x$, $y$.
Here we have written $(\Z^d)^*$ for 
the lattice of bonds of $\Z^d$. Also, the condition 
(2.9) giving continuity of the conditional expectation holds if 
$$
\eqalign{
&\lim_{\L^*\uparrow(\Z^d)^*}
\sup_{\hat J}
\left|\mu_{\infty}[J_{\L^*}
\hat J_{(\Z^d)^*\ba\L^*}](\tilde \s_x
\tilde\s_y)
- \mu_{\infty}[J_{(\Z^d)^*}](\tilde \s_x
\tilde\s_y)\right|=0
}
\tag{5.7}
$$
for nearest neighbors. Finally, the quantity giving 
the decay of the potential is 
$$
\eqalign{
&\bar c(m)\cr
&\leq \Const 
\sup_{{x,y:|x-y|=m}\atop{ e,e'\in \EE}}
\int\P(d J)
\Biggl|\mu[J](\tilde\s_x \tilde\s_{x+e} \tilde\s_y \tilde\s_{y+e'})
-\mu[J](\tilde\s_x \tilde\s_{x+e})
\mu[J](\tilde\s_y \tilde\s_{y+e'})
\Biggr|
\cr
}
\tag{5.8}
$$
for $m$ big enough s.t. $\{x,x+e\}\cap \{y,y+e'\}$ is always 
empty. (Again the sup over the 
possible different choices of $\eta_x$ and $\eta_y$
was absorbed in the constant.)  
This quantity could be called the quenched 
average of the `energy-energy'-
correlation function. 

We expect this to decay faster than any 
power in a very general situation.
Exponential decay of the quantity under the modulus, 
uniformly in $J$ holds of course in a high-temperature 
regime where the maximum of the possible values of 
$|J_{x,e}|$ is sufficiently small. If this value is 
small enough, this can be seen by a usual 
high-temperature cluster expansion. 
This results in the existence of a translation 
invariant potential, whose sup-norm decays 
according to the remark after Theorem 2.4. 

In [K6] we gave a heuristic discussion of the example of a joint measure 
corresponding to a random Dobrushin state 
for a random ferromagnet describing 
a stable interface between the plus and the minus 
state. Such states are believed to exist 
in $d\geq 4$ for low temperature, and weak disorder, 
though this is only proved in the solid-on-solid approximation 
(see [BoK1]).  
We argued that the corresponding joint measure 
should {\bf not} be almost Gibbsian, if the set of 
possible values of the couplings contains a value 
that is small enough such that the corresponding 
homogeneous system is in the high temperature phase. 
Indeed, choosing this coupling in a large annulus 
one can decouple the inside of the system from 
the outside. So, the inside of the system should be 
in a mixture of the ferromagnetic plus resp. minus 
state rather than the Dobrushin state, a difference 
that can be observed on the nearest neighbor correlations. 
Nevertheless, we expect fast decay of the averaged 
correlations (5.8). 
So, as for the random field Ising model in the phase
transition regime, 
we should have another example of a joint measure 
that is not almost Gibbsian, but has a translation-
invariant interaction potential that decays 
faster than any power outside of a set of 
measure zero. 

\bigskip

This following example appears in the physical literature
[Ku1,2], [MKu] and was first rigorously discussed by [EMMS] 
below the percolation threshold.
We are a little more explicit in the discussion than 
in our previous examples. 

\noindent{\thbf (iii) The diluted random ferromagnet (`GriSing field'):} 
The single spin space for the variables $\s_x$ is again 
$\O_0=\{-1,1\}$.
The disorder variables are given by the occupation numbers 
$\eta_x$ taking values in $\{0,1\}$, independently 
w.r.t. $x$ with density $\P[\eta_x=1]=p$. 
The disordered potential $\Phi(\s,\eta)$ is 
given by $\Phi_{\{x,y\}}(\s,\eta)=-J\eta_x\s_x\eta_y\s_y$
for nearest neighbors $x,y\in \Z^d$ and $\Phi_A=0$ else. 
So the one-site variation of the Hamiltonian is  
$\D H_{x}(\s_{\ov x},\eta^1_{x},\eta^2_{x},\eta_{\del {x}})
= -J(\eta^1_{x}-\eta^2_{x})\s_x\sum_{y:d(y,x)=1}\eta_y\s_y$.

By the results of [EMSS] and [K6]
we know that, for {\bf any} $p$, for sufficiently 
large $J$, {\bf any} weak limit of 
the joint measures of the GriSing random field 
is non-Gibbs. [EMSS] noted that, for $p$ below $p_c$, 
the percolation threshold for ordinary site percolation, 
one easily obtains a potential for the joint measure 
by putting $U_A(\eta)=
\log Z^0_{A\ba \del (A^c)}$ for the free energy potential if 
$A\ba \del (A^c)$ is a connected component of 
$\{x,\eta_x=1\}$ and $U_A(\eta)=0$ else. (Here 
$Z^0_{B}$
is the partition function of the ordinary 
fully occupied Ising model on the set $B$ with open 
boundary conditions on $\del B$.) It is 
well-defined on the full-measure set of configurations 
where there is no infinite cluster and (trivially) 
absolutely summable on this set. 

On the other hand, by the general result Theorem 2.1, we know 
that there is a $\P$-normalized potential which 
is convergent for $\P$-a.e. $\eta$ for any value 
of p, $0<p<1$. By Theorem 
2.3 we know that there is a (suitably regrouped) 
potential constructed from this potential that 
converges even absolutely for $\P$-a.e $\eta$.
To be a little more specific:
It is easy to see that in this case a $\P$-normalized 
potential on the disorder space can be written in the 
form $U^{\hbox{\srm{fe}}}_{\mu; A}(\eta)= 
c_A(J,p)\prod_{x\in A}(\eta_x-p)$. 
From the proof of Theorem 2.1 we see that, 
for a given measurable Gibbs measure $\mu[\eta]$, the parameters 
$c_A(J,p)$ of the corresponding  
free energy potential are to be determined from the equations 
(4.3) and (4.4).  A.s. convergence is guaranteed by Theorem 2.1 
and means 
$\sum_{A: A\ni x} 
c_A(J,p)\prod_{y\in A}(\eta_y-p)<\infty $ for $\P$-a.e. $\eta$. 
Note, on the other hand, that we certainly have 
that $\sum_{A: A\ni x} |c_A(J,p)|(1-p)^{|A|}=\infty$ 
for $p\leq \frac{1}{2}$ and 
$\sum_{A: A\ni x} |c_A(J,p)|p^{|A|}=\infty$ 
for $p\geq \frac{1}{2}$ for $J$ sufficiently large.  
This is clear because the above sums are 
just the sums over the sup-norms of the interactions 
and otherwise the potentials would be absolutely uniformly  
summable.

It is however also interesting to discuss the vacuum potentials
and check the hypothesis of Theorem 2.2.  
We start with the potential corresponding 
to the `empty' vacuum $\hat\eta_x^{(0)}\equiv 0$. 
It has the form $V^{\hbox{\srm{fe}}}_{\mu; A}(\eta)= 
c^{(0)}_A(J)\prod_{y\in A}\eta_y$ 
(corresponding to [Ku2(31)]).
Note that the definition of the constants $c^{(0)}_A(J)$
by (4.3) and (4.4) involves only expectations w.r.t. 
$\mu[\hat \eta^{(0)}]$ which is just an infinite product 
over symmetric Bernoulli measures. Trivially, the weak convergence (2.2) 
holds, and is independent of the boundary condition. 
So, the constants are explicitly computable 
up to any desired magnitude of $|A|$. 
In particular, they do not depend on $p$.
Corollary 2 states that, under the hypothesis 
of Theorem 2.2, also the potential of the form 
$c^{(0)}_A(J)\left(\prod_{y\in A}\eta_y -p^{|A|}\right)$
(which corresponds to [Ku2(32)]) 
is an a.s. convergent potential for the joint system. 
The vacuum potential 
with `occupied' vacuum $\hat\eta_x^{(1)}\equiv 1$ has 
the form  $V^{\hbox{\srm{fe}}}_{\mu; A}(\eta)= 
c^{(1)}_A(J)\prod_{y\in A}(\eta_y-1)$. 
By (4.3), (4.4) the constants are expressed in terms of averages w.r.t. 
$\mu[\hat \eta^{(1)}]$  (obtained as weak limit with 
suitably chosen boundary condition.)
We note that these constants 
must be such that $\sum_{A: A\ni x} |c^{(0)}_A(J)|=\infty$ 
and $\sum_{A: A\ni x} |c_A^{(1)}(J)|=\infty$, 
because $\mu[\eta]$ would be a Gibbs-measure else, as above.

\noindent {\bf $p< p_c$} (easy case):
There is a unique quenched Gibbs measure $\P$-a.s.
which is just the independent product over 
the connected components  
of the occupied sites (which are all finite, $\P$-a.s.) . 
Assuming that $\eta$ is such that all 
connected components of occupied sites are finite,  
one has (2.8) for any $\hat \eta$.
From this follows that the vacuum free energy potential 
converges, for any vacuum $\hat \eta$.
In particular one has, for the empty
(resp. the full) vacuum that 
$\sum_{A: x\in A\sb\{y\in\Z^d,\eta_y=1\}} c^{(0)}_A(J)<\infty$ 
(resp. 
$\sum_{A: x\in A\sb\{y\in\Z^d,\eta_y=0\}}(-1)^{|A|}
c^{(1)}_A(J)<\infty$). For the vacuum potential $V_{A}^{(0)}$
with empty vacuum the situation is 
particularly simple: We see by (4.3) and (4.4) that $V_{A}^{(0)}(\eta)=0$ 
unless $A$ is a subset of a connected component 
of $\{x\in \Z^d,\eta_x=1\}$. [Because: (4.3) 
decomposes into a sum 
over the connected components of the occupied sites in $\L$, i.e.
$E_{\L}^{(0)}(\eta)
= \sum_{i}\log Z^0_{B_{\L,i}(\eta)}+C_{\L}$ 
where $B_{\L,i}(\eta)$ are the connected components 
of $\{x\in \L, \eta_x=1\}$ and $C_{\L}$ does not 
depend on $\eta$]. This implies that $c_{A}^{(0)}=0$ 
unless $A$ is connected. 
So, $V^{(0)}_{A}(\eta)$ is just 
obtained by the decomposition of the individual logs of 
partition functions over all subsets $A$ of those connected 
components of occupied sites and is thus a `refinement' of the 
potential given just by the logs.  
Consequently
$\sum_{A:A\ni x}V^{(0)}_{A}(\eta)$ contains only finitely 
many terms for all 
$\eta$ such that $\{y\in \Z^d,\eta_y=1\}$ is finite.

\noindent {\bf $p>p_c$:}
There is an infinite cluster of occupied 
sites with probability one. 
One may have different 
Gibbs measures on this infinite cluster, 
including the ferromagnetic ones, and also, 
in sufficiently high dimensions, low dilution 
and low temperature, Dobrushin type interface 
states (the latter is only partially proved [BoK1]). 

Let us assume at first that $p,J$ are such 
that we have a ferromagnetic plus state 
$\mu^{+}[\eta]$ for $\P$-a.e. $\eta$. We look at the 
vacuum potential with empty vacuum, 
given by the same $p$-independent 
formulas as for the $p<p_c$ case in terms of coupling 
constants $c^{(0)}_{A}$ for connected subsets $A\sb \Z^d$.  
Next we assume that $\eta$ is such that 
the finite volume Gibbs-measures with open 
boundary conditions converge to the symmetric mixture 
$\frac{1}{2}\left(\mu^{+}[\eta]+\mu^{-}[\eta]\right)$. 
But, this means that  
$\mu[\eta_{\L}\hat\eta^{0}_{\Z^d\ba \L}]
\rightarrow \frac{1}{2}\left(\mu^{+}[\eta]+\mu^{-}[\eta]\right)$, 
because, on $\L$, the l.h.s. is nothing but the finite 
volume Gibbs measure with open boundary conditions on 
$\L\cap\{x\in \Z^d,\eta_x=1\}$. 
Thus, the r.h.s. differs from the plus state 
as a measure, so there is {\bf no} continuity 
on the level of measures. However, since the observable 
conjugate to the disorder variables is symmetric 
in $\s$, the corresponding expectations are the same 
for the plus and the minus state and we have (2.8), 
i.e. continuity on the level of the $Q$'s. 
Assuming that the set of $\eta$'s with the 
above property is full measure,  
the vacuum potential converges $\P$-a.s. and 
the corresponding joint potential 
describes the joint measure corresponding 
to the ferromagnetic plus state (and also the minus state). 
Conversely we have 

\proposition{5.1}{\it Consider the dilute Ising ferrogmanet, 
at any fixed $J>0$. Assume that there is a 
convergent free energy vacuum potential with empty vacuum 
$\hat\eta_x=0$ for all $x$ for the joint measure 
corresponding to a given Gibbs-measure $\mu[\eta]$ 
of the form 
$$
\eqalign{
&U^{\hbox{\srm{fe}},0}_A(\eta)
:= c_{A}^{(0)}\prod_{x\in A}\eta_x
}
\tag{5.9}
$$
where $A$ is running over the connected subsets of $\Z^d$. 
Then we must have  
$$
\eqalign{
&c_{A}^{(0)}=\sum_{\L:\L\sb A}(-1)^{|A\ba \L|}
\log \frac{Z_{\L}^0}{2^{|\L|}}\cr
}
\tag{5.10}
$$
where, as above, $Z_{\L}^0$ is the partition 
function of the fully occupied model in $\L$ with 
zero boundary conditions. 
In particular, if two (possibly different) Gibbs-measures 
corresponding to the same $J$ 
both have a potential of the form (5.9), it must be the same. 
}

The proof is given below. Applying the proposition 
to the random Dobrushin (interface) state 
we see that we expect a different scenario for the 
corresponding joint measure. Assuming that there 
is a free energy potential of the form (5.9)
it is the same as for the joint measure of the 
plus state. This is the potential constructed 
from (4.3) in a straightforward way. From (3.1) we see however that the 
conditional expectations in the infinite volume 
will be different in plus-state and Dobrushin-state,
because:  Equality of the l.h.s. of (3.1) for different 
$\mu[\eta]$ implies equality of $Q_x$ for different $\mu[\eta]$ 
(by varying the boundary condition $\xi_{\del x}$). 
The corresponding $Q_x$ in turn 
are essentially given in terms of nearest 
neighbor correlations and these will differ 
in interface states and ordered states. So, both states cannot have 
the same potential. 
This provides an example of a convergent potential 
constructed in a natural way that produces the 
wrong measure. 

Finally we look at the vacuum potential with the fully occupied vacuum. 
We discuss again the joint measure corresponding 
to the ferromagnetic plus state and the Dobrushin state. 
If these states do exist a.s. then they 
also exist for the fully occupied system.  So 
we can construct the state $\mu[\hat\eta]$, and the 
state $\mu[\eta]$ for typical $\eta$ with the 
same type of boundary conditions, in both cases. 
Also, in both cases, we expect 
that $\mu[\eta_{\L}\hat\eta^{1}_{\Z^d\ba \L}]
\rightarrow \mu[\eta]$ which, in particular, implies 
(2.8). So the corresponding vacuum potential 
converges and yields the right conditional 
probabilities. 
Observe, that in a situation where a typical 
realization of the disorder destroys the Dobrushin state
that is present for $\hat\eta^{(1)}$, 
a weak limit of finite volume 
Gibbs measure with plus/minus boundary condition will yield
a symmetric mixture of plus and minus state. 
Thus, to get a correct potential, we should of course 
choose the corresponding 
$\mu[\hat\eta^{(1)}]$ to be (say) the plus state (which yields 
the same free energy potential as the symmetric mixture).  
The Dobrushin state in the ordered system which will result 
from plus/minus boundary conditions will give
a wrong potential. This illustrates the `freedom 
of choice' of the boundary condition for the Gibbs-measure 
with corresponding to $\hat\eta$ offered in Theorem 2.2.

It remains to give the 

\proofof{Proposition 5.1} We claim that in order that the conditional 
expectations be the correct ones we must 
have that 
$$
\eqalign{
&\lim_{\D\uparrow\Z^d}\sum_{A:A\sb \D,A\ni x}
\left(
U^{\hbox{\srm}fe}_A(\eta^1_{x}\eta_{\Z^d\ba x})
-U^{\hbox{\srm}fe}_A(\eta^2_{x}\eta_{\Z^d\ba x})
\right)
= \log Q_{x}(\eta^1_{x},\eta^2_{x},\eta_{\Z^d\ba x})
}
\tag{5.11}
$$
for $\P$-a.e. $\eta$, for all $\eta_x^1$ and $\eta_x^2$.
This 
follows from the fact that the $\D$-limit of (4.2)
(which is assumed to exist) 
and (3.1) must coincide, $\P$-a.e., which 
is equivalent to 
$$
\eqalign{
&\int\mu^{\hbox{\srm ann,}\x_{\del x}}_{x}(d\tilde\eta_{x})
e^{-\sum_{A: A\ni x}
\left(U^{\hbox{\srm}fe}_A(\tilde\eta_{x}\eta_{\Z^d\ba x})
-U^{\hbox{\srm}fe}_A(\eta)
\right)}
= 
\int\mu^{\hbox{\srm ann,}\x_{\del x}}_{x}(d\tilde\eta_{x})
Q_{x}(\eta_{x},\tilde\eta_{x},\eta_{\Z^d\ba x})
}
\tag{5.12}
$$
A simple computation shows that 
the one-site annealed distribution 
is given by  
$\mu^{\hbox{\srm ann,}\x_{\del x}}_{x}(\eta_{x}=1)/
\mu^{\hbox{\srm ann,}\x_{\del x}}_{x}(\eta_{x}=0) 
=\cosh(J\sum_{y\in \del x}\eta_y\s_y)
$. Thus, by writing (5.12) for different values of $\x_{\del \L}$ 
corresponding to different values for the expression in the cosh 
we can conclude that (5.12) really implies (5.11).
Fix $\L$. Knowing that $\mu[\eta]$ satisfies 
the DLR-equation for $\P$-a.e. $\eta$ we have that  
$\mu[\eta_{\L}\hat \eta_{\del \L} 
\eta_{\Z^d\ba \ov{\L}}](\s_{\L})=\mu^0_{\L}[\eta_{\L}](\s_{\L})$, 
for $\P$-a.e. $\eta_{\Z^d\ba \ov{\L}}$. So we have from (5.11)
(putting $\eta^1_x=\eta_x$, $\eta^2_x=\hat \eta_x$)
$$
\eqalign{
&\lim_{\D\uparrow\Z^d}\sum_{A:A\sb \D,A\ni x}
U^{\hbox{\srm}fe,0}_A(\eta_{\L}\hat \eta_{\del \L} \eta_{\Z^d\ba \ov{\L}})
= \log Q_{x}(\eta_{x},\hat\eta_{x},\eta_{\L\ba x}
\hat \eta_{\del \L} \eta_{\Z^d\ba \ov{\L}})
= \log \frac{Z_{\L}^0(\eta_x\eta_{\L\ba x})}{Z_{\L}^0(  
\hat\eta_x\eta_{\L\ba x})}\cr
}
\tag{5.13}
$$
for $\P$-a.e. $\eta_{\Z^d \ba \ov{\L}}$ whenever $x\in \L$.
The l.h.s. equals $\sum_{A:A\sb \L,A\ni x}
U^{\hbox{\srm}fe,0}_A(\eta)$ due to the assumption 
on the form of the potential 
involving only connected $A$'s. From this one sees by 
telescoping over the sites in $\L$ that 
$\sum_{A:A\sb \L}c_{A}^{(0)}=\sum_{A:A\sb \L}
U^{\hbox{\srm}fe,0}_A(1_{A})= \log Z_{\L}^0/2^{|\L|}
$
which, by the inclusion-exclusion formula gives (5.10).\endproof


\ftn
\font\bf=cmbx8

\baselineskip=10pt
\parskip=4pt
\rightskip=0.5truecm
\bigskip\bigskip
\chap{References}
\medskip


\item{[AW]} M.Aizenman, J.Wehr, Rounding Effects of Quenched Randomness
on First-Order Phase Transitions, Comm. Math.Phys {\bf  130},
489-528 (1990)








\item{[BK]} J.Bricmont, A.Kupiainen, 
Phase transition in the 3d random field Ising model,
Comm.
Math.Phys. {\bf 142}, 539-572 (1988)


\item{[BKL]} J.Bricmont, A.Kupiainen, R. Lefevere,
       Renormalization Group Pathologies and the 
Definition of Gibbs States,
Comm. Math.Phys. {\bf 194} 2, 359-388 (1998)
    

\item{[BoK1]} A.Bovier, C.K\"ulske, A rigorous
renormalization group method for interfaces in random
media, Rev.Math.Phys. {\bf 6}, no.3, 413-496 (1994)

\item{[BoK2]} A.Bovier, C.K\"ulske, 
There are no nice interfaces in $2+1$ dimensional 
SOS-models in random media, J.Stat.Phys. {\bf 83}, 751-759
(1996)


\item{[Co]} D.L.Cohn, Measure Theory, Birkh\"auser, 
Boston, Basel, Stuttgart (1980)



\item{[Do1]} R.L.Dobrushin, 
Gibbs states describing a coexistence of phases 
for the three-dimensional Ising model,
Th.Prob. and its Appl. {\bf 17}, 582-600 (1972)

\item{[Do2]} R.L.Dobrushin, 
Lecture given at the workshop `Probability 
and Physics', Renkum, August 1995

\item{[DS]} R.L.Dobrushin, S.B.Shlosman, 
"Non-Gibbsian" states and their Gibbs description, Comm.Math.Phys. 
{\bf 200}, no.1, 125--179 (1999)





\item{[E]} A.C.D.van Enter, The Renormalization-Group peculiarities 
of Griffiths and Pearce: What have we learned?, in: 
Mathematical Results in Statistical Mechanics, Eds S.Miracle-Sol\'e,
J. Ruiz and V. Zagrebnov, (Marseille 1998), World Scientific 1999,
pp.509--526, 
also available as preprint 98-692  at http://www.ma.utexas.edu/mp\_arc

\item{[ES]} A.C.D.van Enter, S.B.Shlosman,
(Almost) Gibbsian description of the sign fields of
SOS fields. J.Stat.Phys. {\bf 92}, no. 3-4, 353--368 (1998)

\item{[EFS]} A.C.D.van Enter, R. Fern\'andez, A.Sokal,
Regularity properties and pathologies of position-space
renormalization-group transformations: Scope
and limitations of Gibbsian theory. J.Stat.Phys.
{\bf 72}, 879-1167 (1993)

\item{[EMMS]} A.C.D.van Enter, C.Maes, 
R.H.Schonmann, S.Shlosman, 
The Griffiths Singularity Random Field,
to appear in the AMS Dobrushin memorial volume,
also available as preprint 98-764  at http://www.ma.utexas.edu/mp\_arc
(1998)

\item{[F]} R. Fernandez, Measures for lattice systems,
Physica A {\bf 263} (Invited papers from Statphys 20, Paris (1998)),
117-130 (1999), 
also available as preprint 98-567 at http://www.ma.utexas.edu/mp\_arc


\item{[Geo]} H.O. Georgii, Gibbs measures and phase transitions, Studies
in mathematics, vol. 9 (de Gruyter, Berlin, New York, 1988)

\item{[Gri]} R.B. Griffiths, Non-analytic behavior above
the critical
point in a random Ising ferromagnet, Phys.Rev.Lett. {\bf 23},
17-20 (1969)


\item{[I]} R.B.Israel,
Convexity in the theory of lattice gases, 
Princeton Series in Physics,
Princeton University Press, Princeton, N.J. (1979)

\item{[K1]} C.K\"ulske, Ph.D. Thesis, Ruhr-Universit\"at Bochum (1993)

\item{[K2]} C.K\"ulske, Metastates in Disordered Mean-Field Models:
Random Field and Hopfield Models,
J.Stat.Phys. {\bf 88} 5/6, 1257-1293 (1997)

\item{[K3]} C.K\"ulske, Limiting behavior of random Gibbs measures: metastates 
in some disordered mean field
models, in: 
Mathematical aspects of spin glasses and neural networks,  
Progr. Probab. {\bf 41}, 151-160,
eds. A.Bovier, P.Picco, Birkh\"auser Boston,
Boston (1998)

\item{[K4]} C.K\"ulske, Metastates in Disordered Mean-Field Models II:
The Superstates,
J.Stat.Phys. {\bf 91} 1/2, 155-176 (1998)

\item{[K5]} C.K\"ulske, A random energy model for size dependence:
recurrence vs. transience, Prob.Theor.
Rel.Fields {\bf 111}, 57-100 (1998)

\item{[K6]} 
C.K\"ulske, (Non-) Gibbsianness and phase transitions in 
random lattice spin models, Mark.Proc.Rel.Fields
{\bf 5}, 357-383 (1999)
, 
preprint available at 
http://www.ma.utexas.edu/mp\_arc/, preprint 99-119 (1999)



\item{[Ku1]} R.K\"uhn, Critical Behavior of the Randomly 
Spin Diluted 2D Ising Model: A Grand ensemble Approach, 
Phys.Rev.Lett. {\bf 73}, No 16, 2268 (1994)

\item{[Ku2]} R.K\"uhn, Equilibrium ensemble Approach to Disordered 
Systems I: General Theory, Exact Results, 
Z.Phys. B {\bf 100}, 231-242 (1996)

\item{[Koz]} O.K.Kozlov, Gibbs Description of a system 
of random variables, Problems Inform. Transmission {\bf 10},
258-265 (1974)


\item{[Le]} R.Lefevere, 
Weakly Gibbsian measures and quasilocality: 
a long-range pair-interaction counterexample, 
J.Stat.Phys. {\bf 95} 3/4, 785--789 (1999)

\item{[MKu]} G.Mazzeo, R.K\"uhn, 
Critical behaviour of the 2d spin diluted 
Ising model via the equilibrium ensemble approach, 
available as cond-mat preprint 9907275 at http://babbage.sissa.it

\item{[Mo]} T.Morita, J.Math.Phys {\bf 5}, 1401 (1964)

\item{[MRM]} C.Maes, F.Redig, A.Van Moffaert, 
     Almost Gibbsian versus Weakly Gibbsian measures,
Stoch.Proc.Appl. {\bf 79}  no. 1, 1--15 (1999), 
also available at 
http://www.ma.utexas.edu/mp\_arc/, preprint 98-193 

Erratum, to appear in Stoch.Proc.Appl.

\item{[MRSM]} C.Maes, F.Redig, S.Shlosman, A.Van Moffaert,
Percolation, Path Large Deviations and Weak Gibbsianity, 
to appear in Comm.Math.Phys., also available as preprint at 
http://www.ma.utexas.edu/mp\_arc/, preprint 99-165 
 


\item{[N]} C.M.Newman, 
Topics in disordered systems, Lectures in Mathematics ETH Zürich.
Birkh\"auser Verlag, Basel, (1997)

\item{[NS1]} C.M.Newman, D.L.Stein, Spatial Inhomogeneity and
thermodynamic chaos, Phys.Rev.Lett. {\bf 76}, No 25, 4821 (1996) 

\item{[NS2]} C.M.Newman, D.L.Stein, Metastate approach to thermodynamic chaos.,  
Phys. Rev. E {\bf 3} 55, no. 5, part A, 5194-5211 (1997)

\item{[NS3]} C.M.Newman, D.L.Stein, 
Simplicity of state and overlap structure in finite-volume realistic
spin glasses, Phys.Rev.E {\bf 3} 57, no. 2, part A, 1356-1366 (1998)


\item{[NS4]} C.M.Newman, D.L.Stein, Thermodynamic chaos and the structure of short-range
spin glasses, in: Mathematical aspects of spin glasses and neural networks, 243-287, 
Progr. Probab., 41, Bovier, Picco (Eds.), Birkh\"auser, 
Boston, Boston, MA  (1998)



\item{[S]} R.H.Schonmann, Projections of Gibbs measures may
be non-Gibbsian, Comm.Math.Phys. {\bf 124}
1-7 (1989)

\item{[Se]} T. Sepp\"al\"ainen, Entropy, limit theorems, 
and variational principles for
disordered lattice systems, Commun.Math.Phys {\bf 171},233-277 (1995)

\item{[Su]} W.G.Sullivan, Potentials for almost 
Markovian Random Fields, Comm.Math.Phys. {\bf 33}
61-74 (1973)

\item{[SW]} G.Sobotta, D.Wagner, Z.Phys. B {\bf 33}, 271 (1979)





\end